\documentclass[final,1p,times]{elsarticle}

\usepackage{graphicx}
\usepackage{amssymb}
\usepackage{array}
\usepackage{amsmath}

\usepackage{tabularx,ragged2e,booktabs,caption}
\newcommand{\SH}{\textsc{$q$-Subset Square Colouring}}
\newcommand{\sh}{subset square colouring}

\newcommand{\EC}{\textsc{X3C}}
\newcommand{\VC}{\textsc{Vertex Cover}}
\newcommand{\col}{\sf col}

\usepackage{array}
\newcolumntype{P}[1]{>{\centering\arraybackslash}p{#1}}
\usepackage{amssymb}
 \newtheorem{theorem}{Theorem}
\newtheorem{lemma}[theorem]{Lemma}
\newtheorem{observation}[theorem]{Observation}
\newdefinition{definition}{Definition}
\newproof{pot}{Proof of Theorem \ref{thm2}}
\newproof{proof}{Proof}

\newtheorem{proposition}[theorem]{\textit{Proposition}}

\newtheorem{corollary}[theorem]{\textit{Corollary}}

\usepackage{lipsum}

\newcommand\blfootnote[1]{%
  \begingroup
  \renewcommand\thefootnote{}\footnote{#1}%
  \addtocounter{footnote}{-1}%
  \endgroup
}

\begin{document}



	\begin{frontmatter}
		
		
		
\title{Colouring a Dominating Set without Conflicts: \\$q$-Subset Square Colouring }

	\author{V P Abidha \fnref{fn1}}
	\ead{abidha.vp@iiitb.ac.in}
	\fntext[fn1]{First author is supported by the TCS Research scholar program.}

%
%
	
		\author{Pradeesha Ashok}
		\ead{pradeesha@iiitb.ac.in}	
		
			\author{Avi Tomar}
		\ead{Avi.Tomar@iiitb.ac.in}

	\author{Dolly Yadav}
		\ead{dolly.yadav@iiitb.ac.in}
		\address{International Institute of Information Technology Bangalore, India}
%
		

	\begin{abstract}
%
%
%

The \emph{Square Colouring} of a graph $G$ refers to colouring of vertices of a graph such that any two distinct vertices which are at distance at most two receive different colours. In this paper, we initiate the study of a related colouring problem called the \emph{subset square colouring} of graphs. Broadly, the \sh{} of a graph studies the square colouring of a dominating set of a graph using $q$ colours. Here, the aim is to optimize the number of colours used. This also generalizes the well-studied Efficient Dominating Set problem. We show that the \SH{} problem is NP-hard for all values of $q$ even on bipartite graphs and chordal graphs. We further study the parameterized complexity of this problem when parameterized by a number of structural parameters. We further show bounds on the number of colours needed to subset square colour some graph classes.
		\end{abstract}
			


		
	\end{frontmatter}
	
\section{Introduction}
\vspace{-5pt}
Vertex colouring is an important problem in the area of graph theory. For a graph $G(V,E)$, the vertex colouring of $G$ refers to a function $f$ from the vertex set $V$ to a set of colours. There are different types of graph colouring problems based on the constraints imposed on this function. A very popular graph colouring question is the Proper Colouring where any two adjacent vertices are to be assigned different colours. Also, several other variants of vertex colouring exist, like harmonious colouring, sigma colouring, metric colouring and acyclic colouring. In addition to the theoretical interest, graph colouring problems are motivated by applications in various fields like register allocation in compilers, job scheduling, transportation networks, etc. See~\cite{chromaticBook} for a detailed reading of graph colouring.
\blfootnote{A primary version of this paper appeared in International Computer Science Symposium in Russia(CSR)2022 \cite{CSR}. }

A number of graph colouring problems are motivated by a problem in Communication called the \emph{Channel Allocation} problem. Here, there exist transmitters $v_1, v_2, \dots, v_n$ and a transmitter may interfere with another transmitter due to a number of reasons. Now the goal is to assign frequencies to the transmitters such that clear reception of signals is guaranteed. This can be represented as a graph where every vertex corresponds to a transmitter and the interference between transmitters is captured by the distance between the corresponding vertices in the graph. Here the frequency assigned to a transmitter corresponds to the colour assigned to the corresponding vertex. In the 90s, Griggs and Yeh~\cite{L} introduced a concept of assigning colours (equivalently, non-negative integers) to vertices such that the assignment of colours to any two vertices depends on whether they are at distance at most two. This is called the $L(h,k)$-colouring of graphs. A colouring $c$ of graph $G$ is an $L(h.k)$-colouring if for any two vertices $u, v  \in V(G)$, $|c(u) -c(v)| \geq h$ if $u$ and $v$ are at distance $1$ and $|c(u) -c(v)| \geq k$ if $u$ and $v$ are at distance $2$. Thus, $L(1,0)$- colouring is the proper colouring itself. Other versions of this problem based on different values of $h$ and $k$ are well-studied~\cite{Lhk}. Note that $L(1,1)$-colouring involves colouring of vertices with non-negative integers such that the colours on adjacent vertices differ by at least $1$ and the colours on vertices at distance $2$ also differ by at least $1$. This graph colouring is also referred to as \emph{Square colouring}~\cite{bu2012optimal,van2003colouring} since it is equivalent to the proper colouring of the square of a graph.


We initiate the study of a variant of Square colouring called \sh. This is defined as follows: 
\begin{definition}
Let $G=(V,E)$ be an undirected graph. A colouring function $c:V(G) \rightarrow \{c_0, c_1,\cdots ,c_q\}$ is called a \emph{$q$-\sh} of $G$ if it satisfies the following constraints:
\begin{itemize}
	\item For every vertex $v$ and every colour $c_i, 1\leq i \leq q$, we have $|c^{-1}(c_i) \cap N[v]| \leq 1$.
	\item A vertex $v$ can have at most $deg(v)$ vertices with colour $c_0$ in $N[v]$, where $deg(v)$ refers to degree of $v$.
\end{itemize}
The open neighbourhood $N(v)$ of the vertex $v$ consists of the set of
vertices adjacent to $v$, i.e., $N(v)=\{u \in V | (u, v) \in E\}$, and the closed neighbourhood of $v$ is $N[v] = \{v\} \cup N(v)$.
\end{definition}
Here, intuitively, assigning the colour $c_0$ to a vertex $v$ corresponds to $v$ being uncoloured. In this paper, we refer to a vertex being uncoloured and a vertex coloured $c_0$, interchangably. Similarly a vertex is said to be coloured if it is coloured using one of the colours from the set $\{c_1 \cdots c_q\}$. Thus the definition implies that every vertex has at least one coloured vertex in its closed neighbourhood and no colour is repeated in the closed neighbourhood. Note that the set of coloured vertices form a dominating set of the graph $G$. Therefore the \sh{} is equivalent to square colouring of a dominating set of the graph using $q$ colours.

  For a given graph $G$, let $\chi_{ssc}(G)$ be the minimum value of $q$ such that there exists a $q$-\sh{} of $G$. We also study the following algorithmic question. Given a graph $G$, the \SH{} problem is defined as follows,
 \vspace{5pt}
 \\\noindent\textbf{Input:} Graph $G$ and $q \in \mathbb{N}$.\\
\textbf{Question:} Can $G$ be $q$-subset square coloured$?$
\vspace{5pt}
\\The concept of subset square colouring is previously studied in the context of a classic problem in Computational Geometry called the Art Gallery problem. Given a polygon $P$, along with two sets, $M$ and $G$, of points in $P$, the Art Gallery problem is to find $G' \subseteq G$ such that every point in $M$ is seen by at least one point in $G'$. Motivated by applications in Robotics, Erickson and LaValle~\cite{erickson2010chromatic} introduced the \emph{Chromatic Art Gallery Problem}. Here, the aim is to find a subset $G' \subseteq G$ such that $G'$ can be coloured using $q$ colours and every point $m \in M$ is seen by at least one point in $G'$ and moreover, every point in $G'$ that sees $m$ gets a distinct colour. It is easy to see that for the case where $M$ and $G$ are the same finite sets, the Chromatic Art Gallery Problem can be reduced to \sh{} of a visibility graph. We believe there exists many other application areas related to Channel Allocation where the \sh{} of graphs becomes useful.

Another motivation for studying the \SH{} problem is that many graphs tend to use much smaller number of colours for \sh{} when compared to number of colours needed for square colouring. For example, complete graphs, star graphs, wheel graphs etc. need $O(n)$ colours for square colouring whereas \sh{} can be done using one colour. This will be useful in many applications where the number of colours corresponds to a resource that needs to be optimized. 

 \noindent We now explore some problems that are related to the \SH{} problem.

\vspace{5pt}
\noindent
\textbf{Related Problems}: The problem of Harmonious colouring was first introduced in 1983 by Hopcroft and Krishnamoorthy \cite{Harmonious} and is defined as follows: The harmonious chromatic number of a graph $G$, denoted by $h(G)$, is the least number of colours which can be assigned to the vertices of $G$ such that each vertex has exactly one colour, adjacent vertices have different colours, and any two edges have different colour pairs. Later, Yue Li Wang et al. \cite{Yue} developed the concept of $d-$ Local Harmonious Chromatic  problem which generalized the Harmonious Chromatic problem. The $d$-Local Harmonious (or just $d$-Harmonious) chromatic problem imposes a restriction that the different colour-pair requirement is only asked to be satisfied for every edge within distance $d$ for any vertex. Thus the $1$-Harmonious chromatic problem is same as the Square colouring problem.


The problem of Efficent Dominating Set~\cite{EDS} for a given graph is also of interest while we study the \sh{} problem. An efficent dominating set is one which is simultaneously an independent and a perfect dominating set. A perfect dominating set $P$ is one in which, for every vertex has exactly one neighbour in its open neighbourhood that belongs to $P$, whereas an independent dominating set $I$ satisfies the condition that set of vertices in $I$ forms an independent set. Specifically, efficent dominating set is a special case of  \sh~with $q=1$. 

Another related problem is the Conflict Free Colouring problem~\cite{ashok2022structural,CFC2} which refers to colouring a dominating set such that every vertex has at least one distinct colour in its closed neighbourhood.



\vspace{-5pt}
\section{A discussion of results}
\vspace{-5pt}
\noindent In this section, we give a summary of our results.

We have already mentioned that the \textsc{Efficient Dominating Set} problem is a special case of \SH. The \textsc{Efficient Dominating Set} problem is known to be $NP$-hard~\cite{EDS}. Thus the \SH{} problem is $NP$-hard for $q=1$. We prove that the \SH{} problem with $q=2$ is $NP$-hard even on planar bipartite graphs and the \SH{} problem is $NP$-hard even on bipartite graphs, for all values of $q$. We also prove that the \SH{} problem is $NP$-hard on chordal graphs. 

We  consider the parameterized complexity of the \SH{} problem. For any problem, an interesting parameter to be studied is the size of the solution. For the \SH{} problem, this parameter will be $q$, the number of colours used. However, this turns out to be much harder than a $W$-hard problem in that it is unlikely to admit an algorithm of running time of the form $f(q)n^{g(q)}$.

\begin{lemma}
\label{qpara}
The \SH{} problem parameterized by $q$ is para-$NP$-hard. 
\end{lemma}
The lemma follows from the fact that the \SH{} problem is $NP$-hard even for $q=1$. 
Moreover, the next result shows that the problem remains $W$-hard even on graphs of diameter $2$.

\begin{theorem}
\label{dia2}
The \SH{} problem parameterized by $q$ is $W[2]-$hard on graphs of diameter $2$.
\end{theorem}
\begin{proof}
For a graph $G$, $\chi_{ssc}(G)$ is bounded by the size of the minimum dominating set, $D$ of $G$. When the diameter of a graph is two, all vertices in any $D$ are at distance at most two from each other. Therefore when we subset square colour the vertices, we can not repeat the colour more than once. This implies that $\chi_{ssc}(G)$ is equal to the size of the minimum dominating set. The minimum dominating set problem is known to be $W[2]$-hard on graphs of diameter $2$ \cite{diameter}. Thus the result follows.
\qed
\end{proof}

With respect to Theorem~\ref{dia2}, we note that the \SH{} problem  is polynomial time solvable on planar graphs with diameter $2$~\cite{PlanarDia}.\\

Next, we consider several structural parameters. A well studied structural parameter for graph problems is the treewidth of the graph. Several hard problems are shown to be $FPT$ when parameterized by treewidth. However, the \SH{} problem can be shown to be $W$-hard when parameterized by treewidth. Next we consider treewidth and number of colours as a combined parameter and show that this is $FPT$. This result is proved using a standard technique in fixed parameter tractable algorithm design called dynamic programming over treewidth.
\begin{figure}[t]
\centering
\includegraphics[scale=0.5]{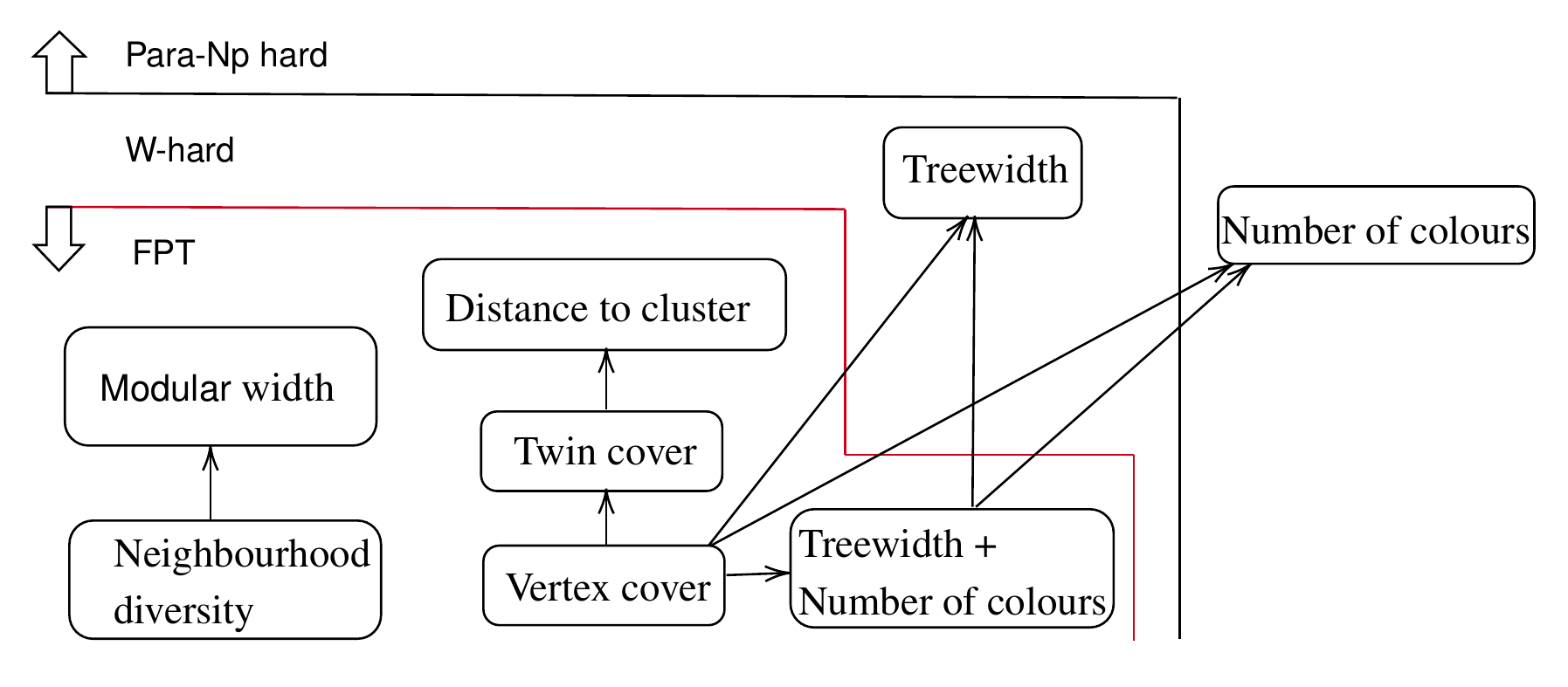}
\caption{Summary of parameterized results.}
\vspace{-5pt}
\label{param_result}
\end{figure}

\noindent Next we consider structural parameters which are possibly larger than treewidth. One such well-studied parameter is the size of the vertex cover of the graph. We give an $FPT$ algorithm for the \SH{} problem parameterized by the size of the vertex cover and we also show that the \SH{} problem does not admit a polynomial kernel when parameterized by the size of the vertex cover unless $NP \subseteq coNP/poly$. The size of vertex cover is usually a large parameter, especially for dense graphs. Therefore, we study a parameter whose value is small on dense graphs called the neighbourhood diversity. It is also a parameter whose value can be computed in polynomial time. We give an $FPT$ algorithm for the \SH{} problem parameterized by neighbourhood diversity and also we show that it admits a polynomial kernel. Further we consider a parameter which is a generalization of neighbourhood diversity, called modular-width and show that the \SH{} problem parameterized by modular-width is $FPT$. We further consider a parameter that is provably smaller than the size of vertex cover, called the \emph{distance to cluster graph} and show that the \SH{} problem parameterized by distance to cluster graph is also $FPT$. We further consider a parameter called \emph{twin cover} whose value typically lies between those of distance to cluster graph and size of vertex cover.
 Since the \SH{} problem parameterized by distance to cluster graph is $FPT$, the \SH{} problem parameterized by twin cover is also $FPT$. However, we give an algorithm with a better running time.

For these problems, as our goal is to only show whether or not the problem is FPT, we do not try to optimize the running times. See Figure~\ref{param_result} for a summary of the results in parameterized complexity of the \SH{} problem.
 
\begin{table}[thb]

\begin{tabular}{|p{6cm}|P{2.5cm}|P{2.5cm}|}
\hline
\textbf{Graph Classes}& \textbf{$\chi_{ssc}$:Upper bound}& \textbf{$\chi_{ssc}$:Lower bound}\\\hline
Path $P_n$ & $1$& $1$\\
\hline
Cycle $C_{3n}$ &$1$& $1$\\
Cycle $C_{3n+1}$ or $C_{3n+2}$& $2$& $2$ \\
\hline
Complete graph $K_n$& $1$& $1$\\
\hline
Complete bipartite graph $G_{n,m}$& $2$& $2$\\
\hline
Bipartite permutation graph $G_{n,m}$& $4$& $3$\\
\hline
Planar graphs with diameter $2$ & $3$& $2$ \\
\hline
Cograph $G_n$& $2$& $2$\\
\hline
Threshold graph $G_n$& $1$& $1$\\
\hline
Caterpillar graph $G(n,r_1,r_2,\cdots r_t)$&$3$& $3$\\
\hline
Tree $T(V,E)$& $O(\Delta)$ & $O(\Delta-1)$\\
Tree $T(V,E)$& $O(n)$ & $\Omega (\sqrt{n})$\\
\hline
Split graph $T(V,E)$& $O(n)$  &$\Omega (\sqrt{n})$\\
\hline 
\end{tabular}
\caption{Summary of bounds for $\chi_{ssc}$ for different graph classes.} 
\label{tab:title} 
\end{table} 
Next we study bounds on $\chi_{ssc}(G)$ when $G$ belongs to certain graph classes. It is easy to see that various graph classes (that include many sparse and dense graph classes) like complete graphs, cluster graphs, star graphs, wheel graphs, paths, cycles, grid graphs etc. are $q$-subset square colourable where $q$ is a constant. However, when we consider trees, we show that there exist trees with $n$ vertices, that requires $O(\sqrt{n})$ colours to be subset square coloured. As trees form a sub-class of bipartite graphs, this result extends for the class of bipartite graphs also. However, we show that a well-defined sub-class of bipartite graphs, called the Bipartite permutation graphs are $4$-subset square colourable. We further show that the class of threshold graphs are $1$-subset square colourable. Note that threshold graphs lie in the intersection of split graphs and cographs. We observe that, while cographs, like threshold graphs, are subset square colourable using a constant number of colours, there exist split graphs which require $O(\sqrt{n})$ colours to be subset square coloured. See Table~\ref{tab:title} for a summary of results.

\section{NP$-$completeness}\label{sectionNPC}
In this section, we show that the \SH{} problem is NP-complete, for all values of $q$. Note that the  \SH{} problem is NP-hard for $q=1$ and it easy to see that the problem is in NP. Therefore \SH{} problem is $NP$-complete when $q=1$. 
Now, we consider $q=2$.

\begin{theorem}
\label{npc2}
	The \SH{} problem, where $q=2$ is NP-complete, even on planar bipartite graphs.
\end{theorem}
\begin{proof}
	 We give a reduction from the planar Exact cover by $3$-sets(X3C) problem.	
\vspace{2pt}  
\\ \noindent\textbf{Planar X3C (Exact cover by $3$-sets) :} \\ \noindent \emph{Input} :  A finite set $X$ with $|X|=3n$ and a collection $S$ of $3$-element subsets of $X$ with $|S| = m$.\\
\emph{Question:} Does $S$ contain an exact cover for $X$, i.e., a sub collection $S' \subseteq S$ such that every element of $X$ occurs in exactly one member of $S'?$
\vspace{4pt}
\\ In Planar \EC{} problem, we have the added constraint that a bipartite graph $M$ such that $V(M)$ corresponds to $X \cup S$ and $E(G)$ is $\{(x,s)|x \in X, s \in S, x \in s\}$ is planar.
\vspace{5pt}
\\Let $(U,S)$ be an instance of the Planar\EC{} problem, where $U= \{ u_1, u_2, .. , u_{3n}\}$ and $\mathcal{S}=\{S_1, S_2, .$ $ . ., S_m \}$.
We construct a planar bipartite graph $G$ as follows:
For every element $u_i$, we add a vertex $x_i$ in $G$ and connect it with an \textit{element gadget} $D_i$ in $G$. For $1 \leq i \leq 3n$, $D_i$ is a tree rooted at a vertex $d_i$, as shown in Figure~\ref{gadget}(a). Each of the two child nodes of $d_i$ are connected to three leaves. 
For every set $S_j$, $1 \leq j \leq m$, we add a \emph{set gadget} $T_j$ with a vertex $t_j$ attached to two leaves $v_j$ and $v_j'$.  We also add a \emph{palette gadget} $P$ which has two vertices $p_1$ and $p_2$ adjacent to each other and each of them attached to three vertices of degree one. See Figure~\ref{gadget}(b). 

\begin{figure}[h]
\begin{minipage}[b]{0.55\linewidth}
\centering
\includegraphics[width=7.3cm]{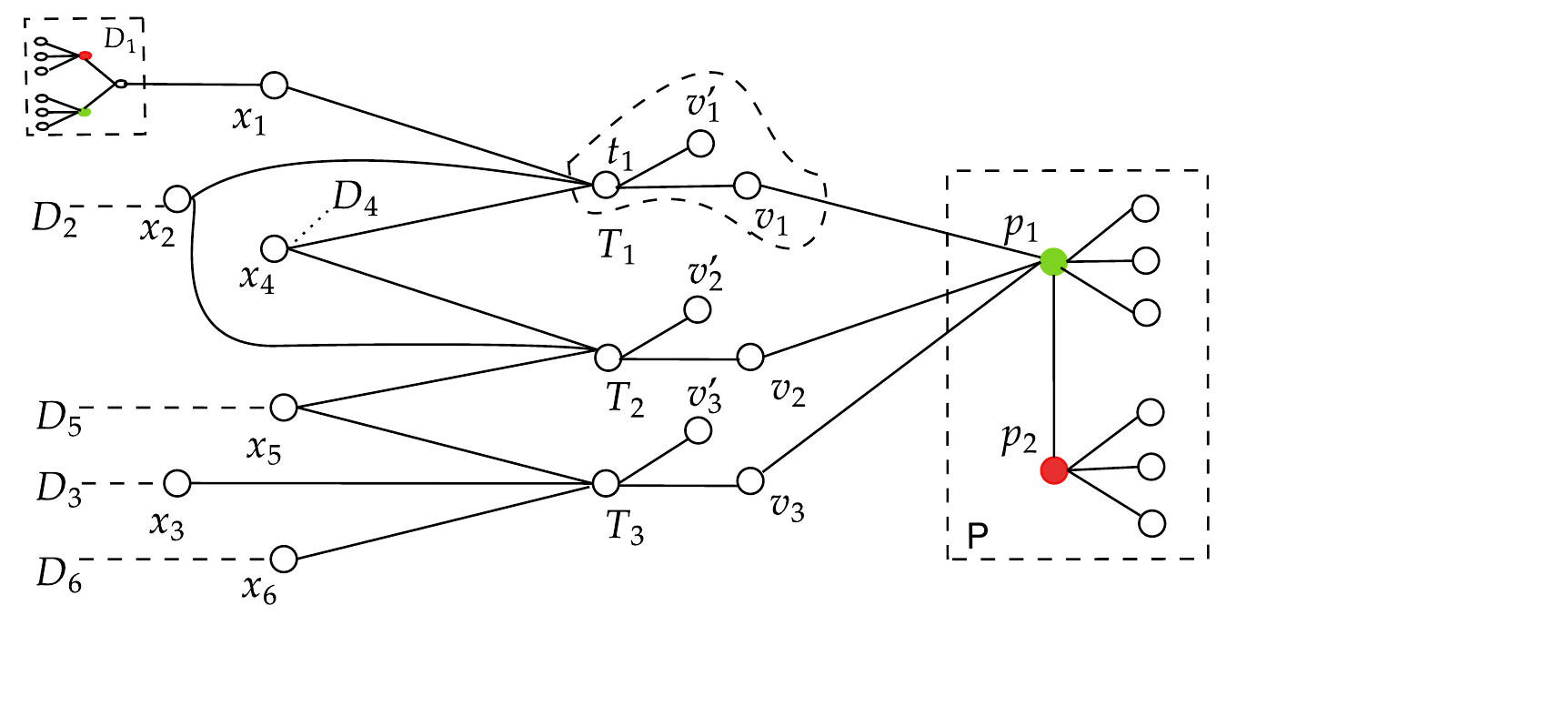}
\caption{Constructed graph $G_{\mathcal{S}}$ from Exact cover by 3-sets insatnce of $S_1=\{x_1,x_2,x_4\}$, $S_2=\{x_2,x_4,x_5\}$ and $S_3=\{x_3,x_5,x_6\}$.}
\label{bpful}
\end{minipage}
\hspace{2cm}
\begin{minipage}[b]{0.24\linewidth}
\centering
\includegraphics[width=2.5cm]{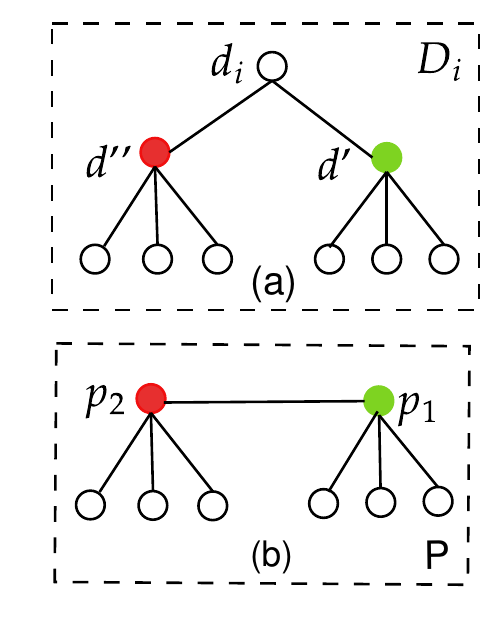}
\caption{(a)Element gadget, (b)Palette gadget}
\label{gadget}
\end{minipage}

\end{figure}

\noindent Further, for $1\leq i \leq 3n, 1 \leq j \leq m$, we add an edge between $x_i$ and $t_j$ in $G$, if $u_i \in S_j$ in $(U,S)$. We also add the edge between $v_j$ and $p_1 \in P$, for all $1\leq j \leq m$. 


We claim that $(U,S)$ has a planar exact-$3$-cover if and only if  $G(V,E)$ has a \sh{} using two colours.

Assume $G$ admits a \sh{}, $\chi$, using two colours. The vertices $p_1$ and $p_2$ will be coloured using different colours since each of them has three neighbours of degree one. Assume, without loss of generality, that $p_1$ is coloured green and $p_2$ is coloured red. Similarly, in every element gadget, the two child nodes of $d_i$ are coloured using different colours. Hence the vertices $d_i$ and $x_i$ cannot be coloured. Therefore $x_i$ need to be dominated by a neighbouring vertex in one of the set gadgets. Moreover, for $1\leq j \leq m$, since $p_1$ is coloured green, $t_j$ which is at distance two from $p_1$, if coloured, needs to be coloured red. Hence, for every vertex $x_i$, there exists a unique $t_j$ which is coloured red and is adjacent to $x_i$. Now it is not difficult to see that $\{S_j | t_j$ is coloured red in $\chi\}$ is a planar exact-$3$-cover for $(U,S)$. See Figure~\ref{bpful}.

%


For the other direction, assume $(U,S)$ has a planar exact-$3$-cover $\mathcal{S}'$. Now we give a \sh{} of $G$ using two colours. Colour the vertices $p_1$ and $p_2$ using green and red colours respectively. For $1 \leq i \leq 3n$, colour the two child nodes of $d_i$ using different colours, in the element gadget $D_i$. For $1 \leq j \leq m$, if $S_j \in \mathcal{S}'$ then colour $t_j$ red, otherwise colour the vertex $v_{j'}$ using any colour. This gives a \sh{} of $G$. 

Now, the result follows from the NP-hardness of the planar \EC{} problem~\cite{E3C}.





\end{proof}
\begin{theorem}
	The \SH{} problem, where $q >2$ is NP-complete, even on bipartite graphs.
\end{theorem}

\begin{figure}[htb]
\centering
\includegraphics[scale=0.6]{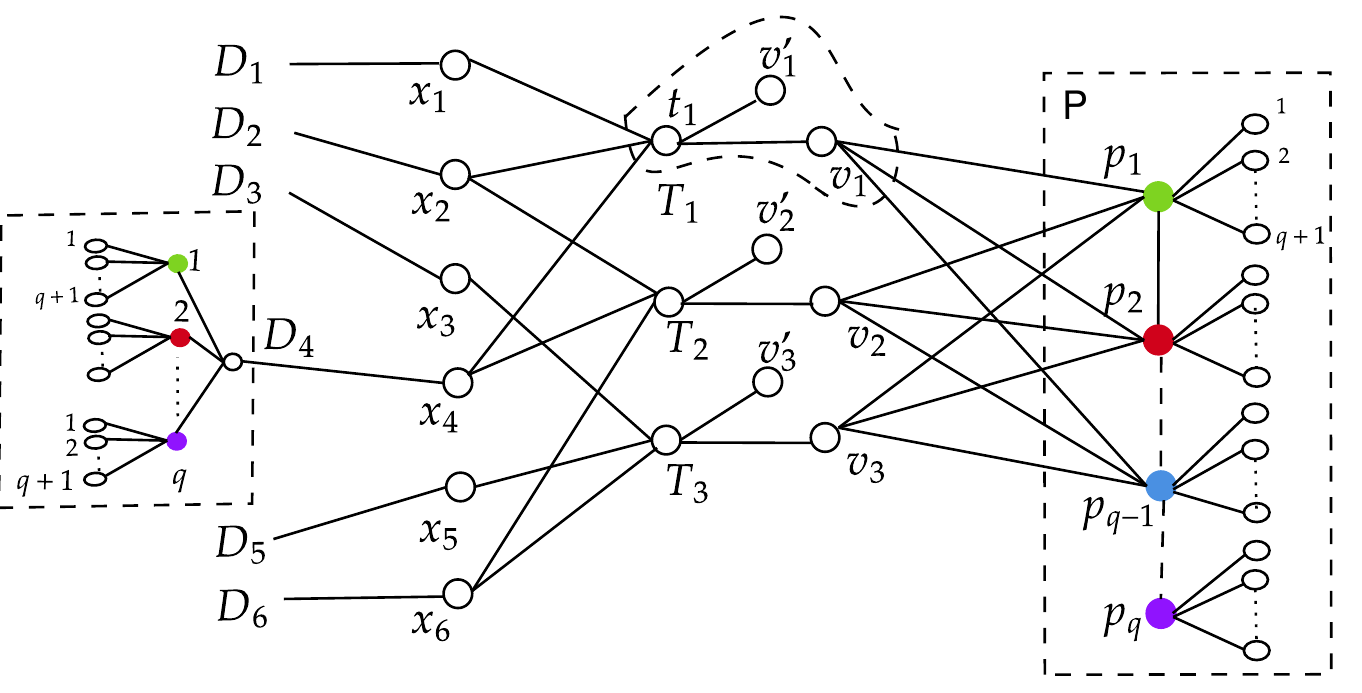}
\caption{Constructed graph $G_{\mathcal{S}}$ from Exact cover by 3-sets insatnce of $S_1=\{x_1,x_2,x_4\}$, $S_2=\{x_2,x_4,x_6\}$ and $S_3=\{x_3,x_5,x_6\}$.}
\label{kfull}
\end{figure}

\begin{proof}
The proof is very similar to that of Theorem \ref{npc2}. We give a reduction from the \EC{} problem. Similar to the previous proof, we construct a bipartite graph $G$ corresponding to an instance $(U,S)$ of the \EC{} problem. Here the element and palette gadgets $D$ and $P$ are slightly different.

For every element $u_i$, we add a vertex $x_i$ in $G$ and connect it with an
\textit{element gadget}, $D_i$, such that $D_i$ is a tree rooted at a vertex $d_i$ and each of the $q$ child nodes of $d_i$ is connected to $q+1$ leaves.
For every set $S_j$, $1 \leq j \leq m$, we add a \emph{set gadget}, $T_j$, in the same way as given in the proof of Theorem~\ref{npc2}. 
We further add a \emph{palette gadget} $P$ which has a path with $q$ vertices $p_1, p_2, \cdots ,p_q$ and each vertex in the path is attached to $q+1$ vertices of degree one. 

Furthermore, for $1\leq i \leq 3n, 1 \leq j \leq m$, we add the edge between $x_i$ and $t_j$ in $G$, if $u_i \in S_j$ in $(U,S)$. We also add edges between $v_j$ and the vertices $p_1, p_2, \cdots ,p_{q-1}\in P$, for all $1\leq j \leq m$. See Figure~\ref{kfull}. 

By the same arguments as given in the proof of Theorem \ref{npc2}, we can now see that $G$ has a \sh{} using $q$ colours if and only if $(U,S)$ has an exact cover by 3-sets.

\end{proof}

\noindent  Now, we show that the \SH{} problem is NP-Complete on chordal graphs for all values of $q$.

\begin{theorem}
\label{npc2}
The \SH{} problem is NP-complete on chordal graphs for all values of $q$. 
\end{theorem}
\begin{proof}
\noindent We give a reduction from the \VC~problem. Let $(G,k)$ be an instance of the \VC~problem, where $|V|=n$ and $|E|=m$. We construct a chordal graph $G' = (V',E')$ as follows. For every vertex $v_i \in V (G)$,  we introduce a vertex $v'_i$ in $G'$ and similarly for every edge $e_{j}\in E(G)$, we introduce a vertex $e'_{j}$ in $G'$.  For every vertex $e'_{j}$ that corresponds to en edge $e_j=(i,h) \in E$, we add edges $(v'_i,e'_j)$ and $(v'_h,e'_j)$. We further connect every pair of vertices $v'_i$ in $G'$, thus forming a clique $C$ of size $|V(G)|$.

Further, for every vertex $e'_{j}$ where $1 \leq j \leq |E(G)|$, we add a subtree $T_j$. The subtree $T_j$ contains a root node $r_j$ with $k$ child nodes and each of these child node is adjacent to $k+1$ leaf nodes. We connect the subtree  $T_j$ with $e'_{j}$ by adding an edge between $r_j $ and $e'_{j}$. 

\noindent 
Apart from the subgraphs that induce trees, the constructed graph $G'$ contains a clique and a set of vertices $e'_{j}$ that are adjacent to vertices of clique. Therefore every induced cycle in $G'$ has  exactly three vertices. Hence, it is easy to see that $G'$ is a chordal graph, and it can be constructed in polynomial time. 
\begin{observation}
\label{Ieaf}
In every $T_j$,  any $k$- subset square colouring colours the $k$ child nodes of $r_j$ using $k$ different colours.
\end{observation} 

The result follows from the fact that the root node $r_j$ is adjacent to $k$ child nodes and each of these child node is adjacent with $k+1$ leaf nodes. Therefore to dominate these leaf nodes, we need to colour the $k$ child nodes of $r_j$ using $k$ different colours.
\begin{figure}[htb]
\centering
\includegraphics[scale=0.6]{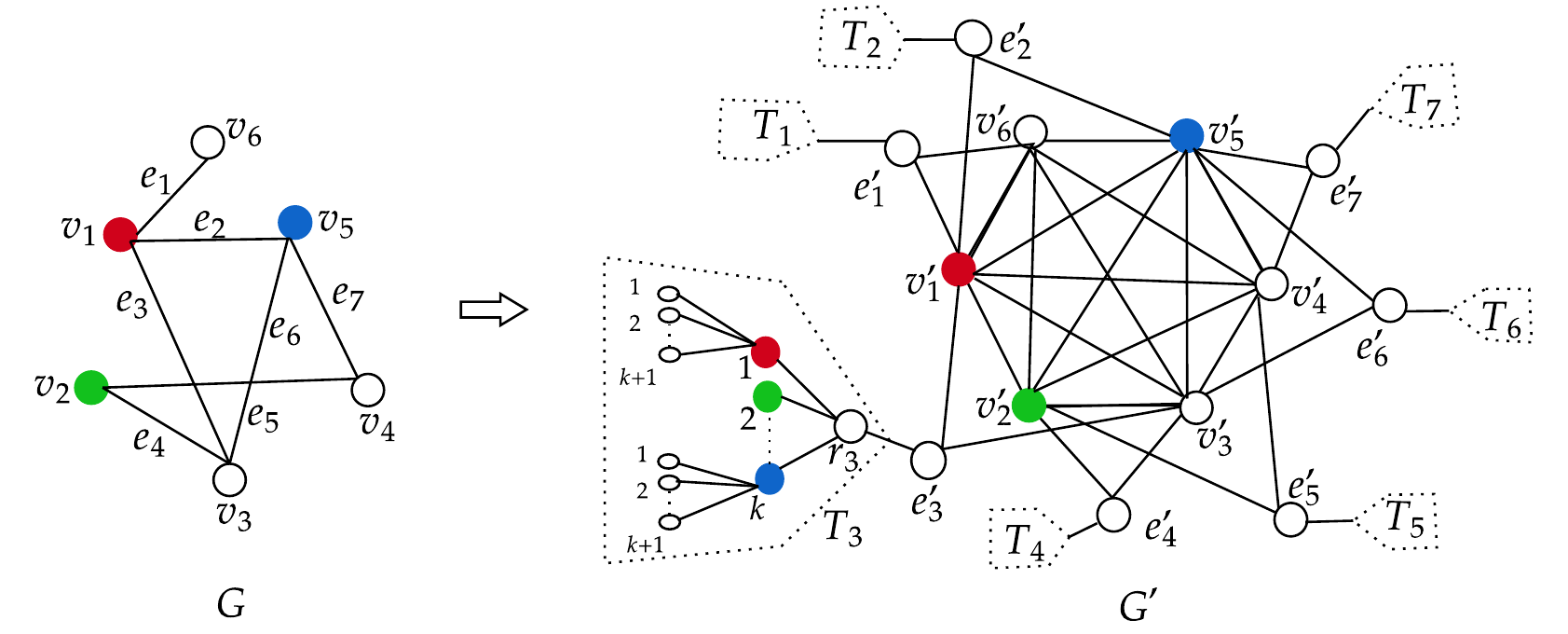}
\caption{Chordal graph $G'(V',E')$ is constructed from the graph $G(V,E)$.}
\label{chordal}
\end{figure} 
\begin{lemma}
The graph $G$ has a vertex cover of size $k$ if and only if  $G'$ admits a $k$-subset square colouring.
\end{lemma}

\begin{proof}
Assume $G'$ has a $k$- subset square colouring. From Observation \ref{Ieaf}, we can see that none of the root nodes in the subtrees and none of the vertices $e'_{j}$ can be coloured, in a $k$- subset square colouring of $G'$. Therefore to dominate the vertex $e'_{j}$, we need to colour either of the vertices in the clique $C$, that it is adjacent to. We can not repeat any colour in $C$ and hence at most $k$ vertices in $C$ are coloured. It is easy to see that the set of coloured vertices corresponds to a vertex cover in $G$. The other direction is similar and easy to see. See Figure \ref{chordal}.

%

\end{proof}
\end{proof}
\noindent Now, the result follows from the NP-completeness of the \VC~problem.
\section{Parameterized Complexity }\label{sectionFPT}
 \noindent
In this section, we study the parameterized complexity of the \SH~problem.

A parameterized problem is a language $L \subseteq \Sigma^* \times \mathbb{N}$, where $\Sigma$ is a fixed, finite alphabet. For an instance $(x, k) \in \Sigma^* \times \mathbb{N}$, $k$ is called the parameter. The complexity class FPT contains all fixed parameter tractable problems that have an algorithm, a computable function $f : \mathbb{N} \rightarrow \mathbb{N}$, and a constant $c$ such that, given $(x, k) \in \Sigma^* \times \mathbb{N}$, the algorithm correctly decides whether $(x, k) \in L$ in time bounded by $f(k)\cdot|(x, k)|^c$ \cite{downey1999parameterized}. Theory of intractability of parameterized problems orders the problems into a hierarchy called the $W$-hierarchy based on its complexity. It is organized as FPT $\subseteq W[1] \subseteq W[2] \cdots $. Under standard complexity theoretical assumptions, a problem which is $W[i]$-hard does not admit FPT algorithms, where $i >0$. For detailed reading of parameterized complexity, refer~\cite{PACygan}.\\

\noindent
As mentioned in the introduction, the \textsc{Efficient Dominating Set} problem is a special case of the \sh{}, when $q=1$. However, the problem of finding an \textsc{Efficient Dominating Set} is NP-hard \cite{EDS}. Hence, the problem of the \SH{} is also NP-hard. Therefore, we explore the possibility of algorithms that are fixed parameter tractable to find \sh{} of a graph. But with respect to the standard parameter $q$, the problem is para-NP-hard since it is NP-hard even for $q=1$ (see Lemma \ref{qpara}). Therefore we investigate the problem of \sh~parameterized by structural parameters. 

\subsection{Parameterized by treewidth}\label{shtree}

\noindent We first consider one of the common and well studied structural parameter of the graph called treewidth. We begin by defining treewidth.\\

\noindent \textbf{Tree Decomposition:} ~\cite{PACygan} A tree decomposition of a graph $G$ is a tree $T$ in
which each vertex $i\in T$ has an assigned set of vertices $X_i \subseteq V$, called the bag, such that $\displaystyle\bigcup_{i\in T} X_i = V$, with some properties:
\begin{itemize}
\item If $u \in X_i$ and $u \in X_k$, then $u \in X_j$ for all $j$ on the path from $i$ to $k$ in $T$. 
\item For any edge $ e(u,v) \in E(G)$, there exists an $i \in T$ such that $u, v \in X_i$.
\end{itemize}

The width of a tree decomposition $T$ is the size of the largest bag of $T$ minus one, and the treewidth of a graph $G$, denoted by $\tau(G)$,  is the minimum width over all possible tree decompositions of $G$.

\noindent Treewidth is a measure of how close a graph is to a tree-like structure. 


\begin{theorem}\label{twW1}
The \SH~problem parameterized by treewidth is $W[1]$-hard.
\end{theorem}

We give a sketch of the proof. A problem which is closely related to the \sh~is the problem of \textsc{L(1,1)-Labeling}. Theorem $5$ of \cite{petr} proves that the \textsc{Square Colouring} problem is $W[1]$-hard parameterized by treewidth. They give a parameterized reduction that convert any instance $(G, r)$ of \textsc{Equitable colouring} problem parameterized by number of colours $r$ to an instance $(H, \tau)$ of \textsc{Square Colouring} problem parameterized by  treewidth $\tau$ such that $\tau$ is bounded by a function of $r$, and the graph $G$ has an equitable colouring by $r$ colours if and only if $H$ has a square colouring using $r$ colours.

\begin{definition}
 \noindent\textbf{Equitable Colouring parameterized by $r$:} \\ \noindent \emph{Input} : A graph $G$ and a positive integer $r$.\\
\emph{Question:} Is there a vertex colouring $c$ of $G$ using at most $r$ colours such that the sizes of any two colour classes differ by at most one $?$
\end{definition}

We modify the reduction given in \cite{petr} as follows to get an instance of the \SH~problem. To the reduced graph $H$, 
we add $q+1$ number of degree-$1$ vertices to each $v \in V(H)$, to obtain the graph $H^\prime$. This forces all the vertices in $V(H)$ to be coloured in any $q$-\sh~of $H'$ since otherwise, if $v \in V(H)$ is not coloured, we have to colour each of the $q+1$ neighbours of $v$ that we introduced in graph $H^\prime$, which clearly overshoots the given budget of $q$ colours. Further, note that this modification does not alter the treewidth of $H$ and treewidth is bounded by a function of $r$. Now, we can see that the graph $G$ has an equitable colouring by $r$ colours if and only if $H'$ has a \sh. Thus, we give a parameterized reduction from the \textsc{Equitable Colouring} problem parameterized by $r$ to the \SH{} problem parameterized by $\tau$. 

\subsection{Parameterized by treewidth and number of colours}

\noindent
We now consider the \SH~problem parameterized by treewidth and number of colours and give an $FPT$ algorithm. We use a standard technique called dynamic programming over treewidth, which gives a constructive proof for the fixed parameter tractability. 
We use a modified tree decomposition called the nice tree decomposition.\\
 
\noindent\textbf{Nice Tree Decomposition:\cite{PACygan}}
A tree decomposition with a distinguished root is called a \textit{nice tree decomposition} if:

\noindent - All leaf nodes and the root node have empty bags, i.e., $X_l = X_r = \phi$, where $r$ is the root node and $l$ is a leaf node.

\noindent - Every other node in the tree decomposition falls in one of the three categories:\\

\noindent \textbf{Introduce node:} An introduce vertex node $t$ has exactly one child $t^\prime$ such that $X_t = X_{t^\prime}\cup\{v\}$ for some $v \not\in X_{t^\prime}$.\\

\noindent \textbf{Forget Node:} A forget node $t$ has exactly one child $t^\prime$ such that $X_t= X_{t^\prime}\setminus \{w\}$ for some $w \in X_{t^\prime}$.\\

\noindent \textbf{Join Node:} A join node $t$ has exactly two children $t_1$ and $t_2$, such that $X_t= X_{t_1}= X_{t_2}$.\\

\noindent\textbf{Introduce edge node:} An introduce edge node is labeled with an edge $uv \in E(G)$ such that $u, v \in X_t$ and has exactly one child node $t^\prime$ such that $X_t = X_{t^\prime}$.
  
Note that we assume every edge is introduced exactly once. 
If a join node contains both $u$ and $v$, and the edge $uv$ exists in $E(G)$, we can note that edge $uv$ will be introduced in the subtree above the join node.

 Nice tree decomposition enables us to add edges and vertices one by one and perform operations accordingly. This variant of tree decomposition still has $O(\tau \cdot n)$ nodes, where $\tau$ is the treewidth of the graph $G$. 

\noindent The following result is known.
\begin{proposition}\label{tw-compute}
Given a graph $G$, in time $2^{O(\tau)}n$, we can compute a nice tree decomposition $(T,{\cal X})$ of $G$ with $|V(T)| \in |V(G)|^{O(1)}$ and of width at most $5\tau$, where $\tau$ is the treewidth of $G$ \cite{5tw}.
\end{proposition}

With each node $t$ of the tree decomposition we associate a subgraph $G_t$ of $G$ defined as: $G_t = \left(V_t, E_t = \{e : e \text{ is introduced in the subtree rooted at t\}} \right)$. Here, $V_t$ is the union of all bags present in the subtree rooted at $t$.
\begin{theorem}\label{thm:StrongCFC}
The \SH~is FPT when parameterized by the treewidth $\tau$ of the input graph and the number of colours $q$.
\end{theorem}

\begin{proof}

\noindent
We give an algorithm based on dynamic programming over nice tree decomposition $(T,{\cal X})$ of $G$, computed in time $2^{O(\tau)}n$, using Proposition~\ref{tw-compute}, of width at most $5\tau$, where $\tau$ is the treewidth of $G$. We define subproblems on $t \in V(T)$ for the graph $G_t$. We consider a \textit{partitioning} of bag $X_t$ by a mapping $f : X_t \rightarrow \{B, W, R\}$. For simplicity, we refer to the vertices in each partition respectively as black, white and grey. Each vertex is also assigned another colour by a function $c$ : $X_{t} \rightarrow \{ c_0,c_1,...,c_q \}$ and a $q$-length tuple, by a function $\Gamma$: $X_{t} \rightarrow \{ 0,1,\hat{1} \}^q$. Roughly speaking, these functions will determine how the ``partial'' square colouring looks like, when restricted to $G_t$ and vertices of $X_t$. Here, $c(v)$ denotes the colour assigned to $v$ and $c(v)=c_0$ denotes that $v$ is not coloured. Also, $\Gamma(v)[i]$ indicates whether $v$ has (either in the current graph, or in the ``future'') a vertex in its closed neighbourhood that has colour $c_i$. Also, $\Gamma(v)[i] = 1$ denotes that vertex $v$ has a vertex in its closed neighbourhood of colour $c_i$ in $G_t$, $\Gamma(v)[i] = \hat{1}$ denotes that vertex $v$ has a vertex in its closed neighbourhood of colour $c_i$, that is not present in $G_{t}$, but will appear in the ``future'', and $\Gamma(v)[i] = 0$ denotes the absence of colour $c_i$ in the closed neighbourhood of $v$. We slightly abuse the notation and use $\Gamma(v)[c_i]$ and $\Gamma(v)[i]$ interchangeably. In the following we give a detailed insight into the functions $f$, $c$ and $\Gamma$.\\
\noindent \textbf{Black}, represented by $B$. Every black vertex $v$ is given a colour $c(v) \neq c_0$ in a \sh.\\
\noindent\textbf{Grey}, represented by $R$. A grey vertex $v$ is not coloured, not dominated, i.e. $c(v) = c_0$ and for each $i \in [q]$, it has $\Gamma(v)[i] \in \{0,\hat 1\}$. \\
\noindent\textbf{White}, represented by $W$. A vertex $v$ that is neither black nor grey is a white vertex. Note that for a white vertex $v$, $c(v) =c_0$ and there is $i \in [q]$, such that $\Gamma(v)[i]=1$.

\noindent A tuple $(t,c,\Gamma,f)$ is \emph{valid} if the following conditions hold for every vertex $v \in X_t$:

\begin{enumerate}
\item $f(v)=B \implies c(v) \ne c_0$ and $\Gamma(v)[c(v)] = 1$,
\item $f(v)=R \implies c(v)=c_0$ and $\Gamma(v)[i] \in \{0,\hat{1}\}$, $\forall i \in \{1...q\}$, and 
\item $f(v)=W \implies c(v)=c_0$ and $\Gamma(v)[i] = 1$ for some $i \in \{1...q\}$.
\vspace{1pt}
\end{enumerate}

\noindent For a node $t\in V(T)$, for each valid tuple $(t,c,\Gamma,f)$, we have a table entry denoted by $D[t,c,\Gamma, f]$. We have $D[t,c,\Gamma, f]= true$ if and only if there is ${\sf col} : V_t \rightarrow \{c_0,c_1,\ldots, c_q\}$ (where $c_0$ denotes no colour assignment), such that:

\begin{enumerate}
\item ${\sf col}|_{X_t} = c$,

\item for each $v \in X_t$ and $i\in \{1,2,\dots q\}$ with $\Gamma(v)[i] = 1$, there is exactly one vertex $u\in N_{G_t}[v]$, such that $\col(u) = c_i$,

\item for each $v \in X_t$ and $i\in [q]$ with $\Gamma(v)[i] \in \{0,\hat 1\}$, there is no vertex $u \in N_{G_t}[v]$, such that $\col(u) = c_i$, and

\item for each $v\in V_t \setminus X_t$, there is at least one vertex $u\in N_{G_t}[v]$, such that $\col(u) \neq c_0$, and for all such vertices $u$, every other $u' \in N_{G_t}[v]$ have ${\sf col} (u^\prime) \neq {\sf col}(u)$.
\end{enumerate}
\noindent
In the above, such a colouring $\col$ is called a $(t,c,\Gamma,f)$-\emph{good} colouring. (At any point of time wherever we query an invalid tuple, then its value is $false$ by default.) 
 Note that $D[r,\emptyset,\emptyset,\emptyset] = true$, where $r$ is the root of the tree decomposition, if and only if $G$ admits a \sh~using (at most) $q$ colours. \\ 
We define $f_{v\rightarrow \gamma} $ where $\gamma \in \{B,W,R\}$, as the function where $f_{v\rightarrow \gamma}(x) = f(x)$, if $x\neq v$, and $f_{v\rightarrow \gamma}(x) = \gamma$, otherwise. Similarly, we define the functions $c_{v\rightarrow c_i}$ and $\Gamma_{v[i]\rightarrow \alpha}$ where $\alpha \in \{0,\hat 1,1\}$ .
\noindent We now proceed to define the recursive formulas for the values of $D$. \\

\noindent\textbf{Leaf node.} For a leaf node $t$, we have $X_t = \emptyset$. Hence, the only entry is $D[t, \emptyset,\emptyset,\emptyset]$. Moreover, by definition, we have $D[t, \emptyset,\emptyset,\emptyset] = true$.\\

\noindent\textbf{Introduce vertex node.} Let $t$ be the introduce vertex node with a child $t'$ such that $X_t = X_{t'} \cup \{v\}$ for some $v \not\in X_{t'}$. Since the vertex $v$ is isolated in $G_t$, the following recurrence follows.
\begin{align*}
    D[t, c, \Gamma, f] = &
    \begin{cases}
    D[t', c_{|X'}, \Gamma_{|X'}, f_{|X'}] \text{ if } & f(v) = B \text{ and } \Gamma(v)[c_i] \in \{0,\hat{1}\} \text{ }, \forall \text{ } c_i \neq c(v)\\
    D[t', c_{|X'}, \Gamma_{|X'}, f_{|X'}] \text{ if } & f(v) = R \\
    \text{False } & \text{otherwise}
    \end{cases}
\end{align*}

\noindent\textbf{Introduce edge node.} Let $t$ be an introduce edge node labeled with an edge $u^*v^*$ and let $t'$ be the child of it. Thus $G_{t'}$ does not have the edge $u^*v^*$ but $G_t$ has. Consider distinct $u,v\in \{u^*,v^*\}$. 
\begin{enumerate}
\item If $f(u)=B$, $f(v)=W$ and $\Gamma(v)[c(u)] = 1$. We set $D[t, c, \Gamma, f]= D[t', c,$ $ \Gamma_{v[c(u)]\rightarrow \hat{1}}$, $f_{v\rightarrow R}] \vee D[t', c, \Gamma_{v[c(u)]\rightarrow \hat{1}}$, $f_{v\rightarrow W}]$ (if any of the entries are invalid, then it is $false$). 
\item If $f(u)=f(v)=B$ and $ \Gamma(v[c(u)]) = \Gamma(u[c(v)]) = 1$, set $D[t, c, \Gamma, f]= D[t', c, \Gamma_{v[c(u)]\rightarrow\hat{1},u[c(v)]\rightarrow\hat{1}}, f] $.
\item If $\{f(u),f(v)\} \cap \{B\} = \emptyset$, then $D[t, c, \Gamma, f]=D[t', c, \Gamma, f]$.
\item If none of the above conditions hold then $D[t, c, \Gamma, f]=false$.
\end{enumerate}
\begin{lemma}
Recurrence for introduce edge node is correct.
\end{lemma}
\begin{proof}
Note that all the vertices except $u$ and $v$ in $X_t$ are unaffected by introduction of edge $u^*v^*$. Also if neither $u$ nor $v$ are black, then the edge $uv$ cannot affect the domination of any vertex. Therefore the value of $D$ for that tuple is same as that in the child node.

Consider the case where $f(u) =B$ and $f(v)=W$ and $\Gamma(v)[c(u)] = 1$. We show that $D[t,c,\Gamma,f]=true$ when at least one of $D[t^\prime,c,\Gamma_{v[c(u)]\rightarrow \hat{1}},f_{v\rightarrow R}]$ and $D[t^\prime,c,\Gamma_{v[c(u)]\rightarrow \hat{1}},f]$ is $true$. Let $D[t^\prime,c,\Gamma_{v[c(u)]\rightarrow \hat{1}},f_{v\rightarrow R}]=true$. Let $\col : V_{t\prime} \rightarrow \{c_0,\dots,c_q\}$ be a $(t^\prime,c,\Gamma_{v[c(u)]\rightarrow \hat{1}},f_{v\rightarrow R})$-good colouring in $G_{t^\prime}$. Note that $\col(v) =c_0$ and there is no vertex $v' \in N_{G_{t'}}[v]$ such that $\col(v')=c(u)$. Therefore, $\col$ is also a $(t,c,\Gamma,f)$-good colouring since $u$ is the unique vertex in $N_{G_t}[v]$ with $\col(u)=c(u)$. 
 Hence $D[t,c,\Gamma,f]=true$. Similarly, we can prove that when $D[t^\prime,c,\Gamma_{v[c(u)]\rightarrow \hat{1}},f]=true$, any $(t^\prime,c,\Gamma_{v[c(u)]\rightarrow \hat{1}},f)$-good colouring is also a $(t,c,\Gamma,f)$- colouring. In the reverse direction, assume $D[t,c,\Gamma,f]=true$ and let $\col : V_{t} \rightarrow \{c_0,\dots,c_q\}$ be a $(t,c,\Gamma,f)$ - good colouring. Therefore, $u$ is the unique vertex in $N_{G_t}[v]$ such that $\col(u)=c(u)$. If there exists a $c_i \ne c(u)$ such that $v$ has a neighbour $v'$ in $N_{G_t}[v]$ with $\col(v')=c_i$, then $\col$ is a $(t^\prime,c,\Gamma_{v[c(u)]\rightarrow \hat{1}},f)$-colouring, otherwise $\col$ is a $(t^\prime,c,\Gamma_{v[c(u)]\rightarrow \hat{1}},f_{v\rightarrow R})$-colouring. We can prove the correctness for other cases by similar arguments.
\end{proof}

\noindent \textbf{Forget node.} Let $t$ be a forget node with child $t'$ such that $X_t = X_{t'} \setminus \{v\}$ for some $v \in X_{t'}$. Since the vertex $v$ does not appear again in any bag of a node above $t$, $v$ must be either black or white (otherwise, we set the entry to $false$). 
\begin{align*}
    D[t, c, \Gamma, f] =  \bigvee_{\substack{1 \leq i \leq q \\ \alpha \in \{0,1\}^q}} \left( D[t', c_{v\rightarrow c_0}, \Gamma_{v\rightarrow\alpha}, f_{v\rightarrow W}]   \vee  D[t', c_{v\rightarrow c_i}, \Gamma_{v\rightarrow\alpha}, f_{v\rightarrow B}] \right)
\end{align*}

\noindent\textbf{Join node.} Let us denote the the join node by $t$. Let $t_1$ and $t_2$ be the children of $t$. We know that $X_t = X_{t_1} = X_{t_2}$ and $X_t$ induces an independent set in the graphs $G_t$, $G_{t_1}$ and $G_{t_2}$. We say that the pair of tuples $[t_1,f_1,c_1,\Gamma_1]$ and $[t_2,f_2,c_2,\Gamma_2]$ are $[t,f,c,\Gamma]$-consistent if for every $v \in X_t$ the following conditions hold.
\begin{itemize}
\item  If $f(v) = B$ then $(f_1(v), f_2(v)) = (B,B)$ and $c_1(v) = c_2(v)=c(v)$. 
\item  If $f(v)=W$ then $(f_1(v), f_2(v)) \in \{(W,R),(R,W),(W,W)\}$.
\item  If $f(v) = R$ then $(f_1(v),f_2(v))=(R,R)$. 
\item  If $\Gamma(v)[i] = 0$ then $(\Gamma_1(v)[i], \Gamma_2(v)[i]) \in \{(0,0)\}$ for $1 \leq i \leq q$.
\item  If $\Gamma(v)[i] = 1$ then $(\Gamma_1(v)[i], \Gamma_2(v)[i]) \in \{(1,\hat{1}),(\hat{1},1)$\} for $1 \leq i \leq q$.
\item   If $\Gamma(v)[i] = \hat{1}$ then $(\Gamma_1(v)[i], \Gamma_2(v)[i]) \in \{(\hat{1},\hat{1})\}$ for $1 \leq i \leq q$. 
\end{itemize}

It is easy to see that a vertex $v$ belongs to Black partition in $X_t$ if it is Black in $X_{t_1}$ and $X_{t_2}$ and the colour $c(v)$ that is assigned to $v$ is same in these bags. Similarly, we can understand for a vertex in Grey partition. If vertex $v$ belongs to White partition, it implies that it is dominated in exactly one of the bags $X_{t_1}$ and $X_{t_2}$ or both. However if $v$ is dominated in both the bags, then the colour that dominates it is different which is taken care by $\Gamma(v)[i]$. The value of $\Gamma(v)[i]$ is set to 1 for different values of the $q$-length tuple. Therefore, for a vertex to belong to White partition requires one of the three combinations: $\{(W,R),(R,W),(W,W)\}$. Now consider the values that $\Gamma(v)[i]$ can take in $X_t$. If colour $i$ is not present in $N[v]$ in any of the child nodes $t_1$ and $t_2$, then it will naturally not be present in $N[v]$ in $G_t$, implying $\Gamma(v)[i] = 0$. However, $\Gamma(v)[i] = 1$ indicates the presence of a vertex with colour $i$ in $N[v]$ in $G_t$. We further note that colour $i$ appears in exactly one of the child nodes $X_{t_1}$ or $X_{t_2}$ but not both because presence of colour $i$ in both child nodes implies the presence of two vertices with colour $i$ in $N[v]$ in $G_t$. This follows from the observation that $X_t$ induces an independent set in $G_t$, by the property of nice tree decomposition. The next possibility of $\Gamma(v)[i]$ is $\hat{1}$. For this to be true, the colour $i$ should not be present in $G_t$ seen so far but will appear in the tree decomposition eventually.\\

\noindent We set $ D[t, c, \Gamma, f] = \bigvee_{\substack{(f_1,f_2)}}\left({ D[t_1, c_1, \Gamma_1, f_1] \wedge D[t_2, c_2, \Gamma_2, f_2]} \right)$,  where $[t_1,f_1$, $c_1,\Gamma_1]$ and \\$[t_2,f_2,c_2,\Gamma_2]$ is $[t,f,c,\Gamma]$-consistent.\\

We have described the recursive formulas for the values of $D[\cdot]$. Note that we can compute each entry in time bounded by $2^{O(q\tau)}\cdot  q^{O(\tau)} n^{O(1)}$. Moreover, the number of (valid) entries for a node $t\in V(T)$ is bounded by $2^{O(q\tau)}\cdot q^{O(\tau)} n^{O(1)}$, and $V(T) \in n^{O(1)}$. Thus we can obtain that the overall running time of the algorithm is bounded by $2^{O(q\tau)} n^{O(1)}$. 
\end{proof}

\subsection{Parameterized by the size of vertex cover}\label{sharVC}

In this section, we prove that the \SH~parameterized by the size of vertex cover is FPT.  Let $X$ be a vertex cover of $G$, $|X| =k$. First we prove the following result.
\begin{lemma}\label{sharVCbound}
The number of colours required to subset square colour the vertices of graph $G$ is at most the size of vertex cover of $G$.
\end{lemma}

\begin{proof}
We know that every edge $e \in E(G)$ has at least one of its end vertices in vertex cover of $G$. Consider a graph $G$ with a vertex cover $X$ of size $k$. Firstly, assign one of the $k$ colours to all isolated vertices, if they exist. Next colour every vertex in $X$ using $k$ different colours. 
Since we colour each vertex of $X$, every vertex $v \in V$ is either coloured or has at least one coloured vertex in its open neighbourhood. Moreover, each vertex in $X$ is coloured using a different colour which avoids overlapping of colours. 
The result follows.
\end{proof}

Since both the treewidth and number of colours required to subset square colour is bounded by $|X|$, the following theorem follows from Theorem~\ref{thm:StrongCFC}.

\begin{theorem}\label{vertexcoverFPT}
The \SH~parameterized by the size of vertex cover is FPT.
\end{theorem}
Now, we discuss the existence of polynomial kernels for \SH~problem parameterized by the size of vertex cover. A polynomial kernel exists for the \SH~problem, when $q=1$ by a result given in \cite{ashok2022polynomial}.
We prove that the \SH~problem parameterized by the size of vertex cover does not have a polynomial kernel unless $NP \subseteq coNP/poly$ for $q \geq 2$. Here we use the or-cross-composition technique for proving kernelization lower bounds. In order to simplify the proof, we first prove the result for $q=2$, and then we extend the result for $q \geq 3$.

\begin{definition}
\textit {\footnotesize OR}-Cross-composition~\cite{PACygan}. Let  $L \subseteq \Sigma^*$ be a language and $Q \subseteq \Sigma ^* \times N$ be a parameterized language. We say that $L$ cross-composes into $Q$ if there exists a polynomial equivalence relation $\textit{R}$ and an algorithm $\mathcal{A}$, called the cross-composition, satisfying the following conditions. The algorithm $\mathcal{A}$ takes as input a sequence of strings  $x_1, x_2, \cdots, x_t \in \Sigma^*$ that are equivalent with respect to $\textit{R}$, runs in time polynomial in $\sum_{i=1}^{t} |x_i|$, outputs one instance $(y, k) \in \Sigma ^* \times N$ such that:

\begin{itemize}
\item $k \leq p(\max_{i=1}^t |x_i| + \log t)$ for some polynomial $p(.)$, and
\item $(y, k) \in Q$ if and only if there exist at least one index $i$ such that $x_i \in L$.
\end{itemize}
\end{definition}

\noindent We use the technique {\footnotesize OR}-cross-composition given by Bodlaender et al.\cite{bodlaender2014kernelization}. 
\begin{definition}
 If an NP-hard problem $L$, {\footnotesize OR}-cross-composes into a parameterized problem $Q$, then Q does not admit a polynomial kernel unless $NP \subseteq coNP/poly$ and the polynomial hierarchy collapses \cite{bodlaender2014kernelization}.
\end{definition}

\subsubsection{Kernalization: $2$-\textsc{Subset square colouring} problem parameterized by the size of vertex cover}
\label{2nopoly}
We show that $2$-\textsc{Subset square colouring} problem parameterized by the size of vertex cover does not admit a polynomial kernel unless $NP \subseteq coNP/poly$ by giving an {\footnotesize OR}-cross-composition from the NP-hard problem \textsc{Clique}. This framework is similar to the framework used in \cite{bodlaender2021parameterized, ashok2022structural}.
 
In this reduction we consider $t$ instances $X_1,X_2, \cdots ,X_t$ of the \textsc{Clique}.
The instances are said to be equivalent if the number of vertices of each $X_i$ is $n$ and asks for a $k$ sized clique. 
Let  $\{v_{l1},v_{l2}, \cdots, v_{ln}\}$ be the set of vertices in an arbitrary instance $X_l$, where $1 \leq l \leq t$. In this section, we assume that $c_1=$ red and $c_2=$ blue.





\vspace{.1cm}
\noindent We define a \emph{special vertex} before creating an instance $G$ for $2$-\textsc{Subset square colouring}.\\

\begin{figure}[tbh]
\centering
\includegraphics[width=8.5cm]{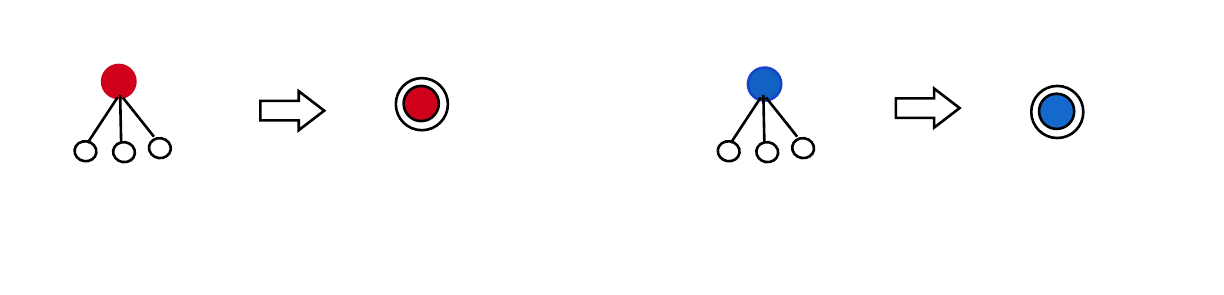}
\caption{Special vertex coloured with red and blue respectively.}
\label{special}
\end{figure}

\noindent \textbf{Special vertex :} If $N(v)$ contains three single-degree vertices, then the vertex $v$ is a special vertex. We denote the special vertex by two circles with the same centre and different radius (Figure \ref{special}). 

\begin{observation}
\label{dum}
For any $2$-\sh~$C$ of $G$, if a vertex $v$ in $G$ is a special vertex, then $C(v) \neq c_0$. 
\end{observation}

We follow some steps to create an instance $G$ of the $2$-\textsc{\sh}. We add two special vertices $R$ and $B$ and we add an edge between $R$ and $B$. This ensures that the vertices $R$ and $B$ receive distinct colours in any $2$-\sh{}. Without loss of generality, assume that $R$ receives the colour red and $B$ receives the colour blue.

%

Now we add a set of vertices $\{ p_{i,j}|i \in [k], j \in [n] \}$, called grid vertices of $G$. These vertices are intuitively used to select the vertices that correspond to a clique in one of the input instances.


In order to complete the construction, we need to create the following gadgets. We connect those gadgets to the grid vertices and also with the special vertices $R$ and $B$.

\begin{figure}[th]
	\begin{minipage}[b]{0.3\linewidth}
		\centering
		\includegraphics[width=4.5cm]{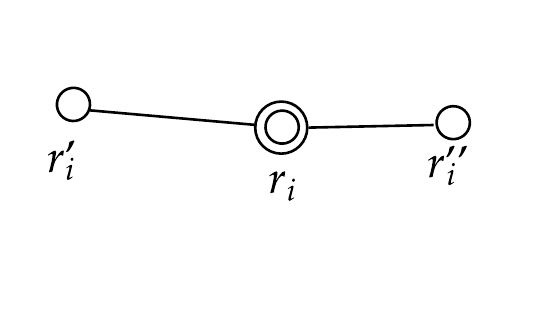}
		\caption{Path gadget $\mathcal{P}1_i$}
		\label{path}
	\end{minipage}
	\begin{minipage}[b]{0.3\linewidth}
		\centering
		\includegraphics[width=3.5cm]{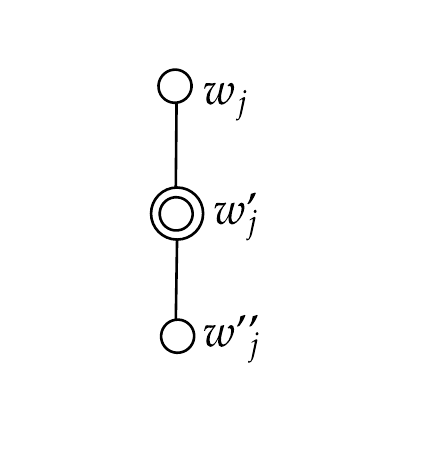}
		\caption{Path gadget $\mathcal{P}2_j$}
		\label{tree}
	\end{minipage}
	\begin{minipage}[b]{0.3\linewidth}
		\centering
	\includegraphics[width=5.5cm]{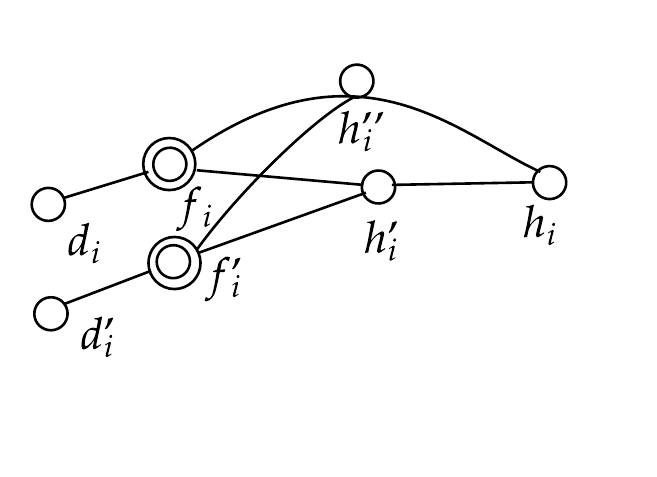}
\caption{Hanging gadget}
	\label{hhang}
	\end{minipage}
\end{figure}

For every $i \in [k]$,  we add a path gadget $\mathcal{P}1_i$ with three vertices $r'_i, r_i,$ and $ r''_i$. In every $\mathcal{P}1_i$, there exist edges $(r'_i, r_i)$ and $(r_i, r''_i)$, where $r_i$ is a special vertex. These vertices form a path in the path gadget. See Figure \ref{path}. In addition to that, connect $\mathcal{P}1_i$ to the vertex $R$, by adding an edge between $r'_i$ and $R$. Moreover $\mathcal{P}1_i$ connects to grid vertices through $r_i$ by adding edges $(r_i, p_{i,j})$, for all $j\in [n]$. 

For every $j \in [n]$, we use a similar path gadget, $\mathcal{P}2_j$ with three vertices $w_j$, $w'_j$ and $w''_j$. In every $\mathcal{P}2_j$, there exist edges $(w_j, w'_j)$ and $(w'_j, w''_j)$, where $w'_j$ is a special vertex. See Figure \ref{tree}. Moreover, we add a vertex $w$ and connect $\mathcal{P}2_j$ to the vertex $R$, by adding edges $(w_j, w)$ and $(w, R)$. Also, connect the vertices $w''_j$ and $R$. In addition to that, we connect $\mathcal{P}2_j$ with grid vertices by adding edges $(w_j,p_{i,j})$, for all $i \in [k]$.

\begin{observation}
\label{pathsSD}
For any $2$-\sh~$C$ of $G$, for all $i \in [k], j \in [n]$, $C(r_i)=$ blue and $C(w'_j)=$ blue and all other vertices in path gadgets $\mathcal{P}1_i$ and $\mathcal{P}2_j$ are dominated by $r_i$ and $w'_j$, respectively.
\end{observation}	

\begin{figure}[thb]
	\begin{minipage}[b]{0.53\linewidth}
		\centering
		\includegraphics[width=5cm]{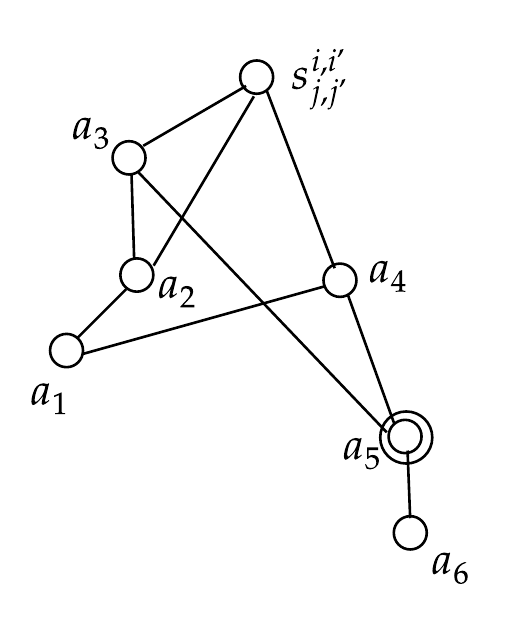}
		\caption{Connector gadget}
		\label{con}
	\end{minipage}
	\begin{minipage}[b]{0.38\linewidth}
		\centering
\includegraphics[width=4.5cm]{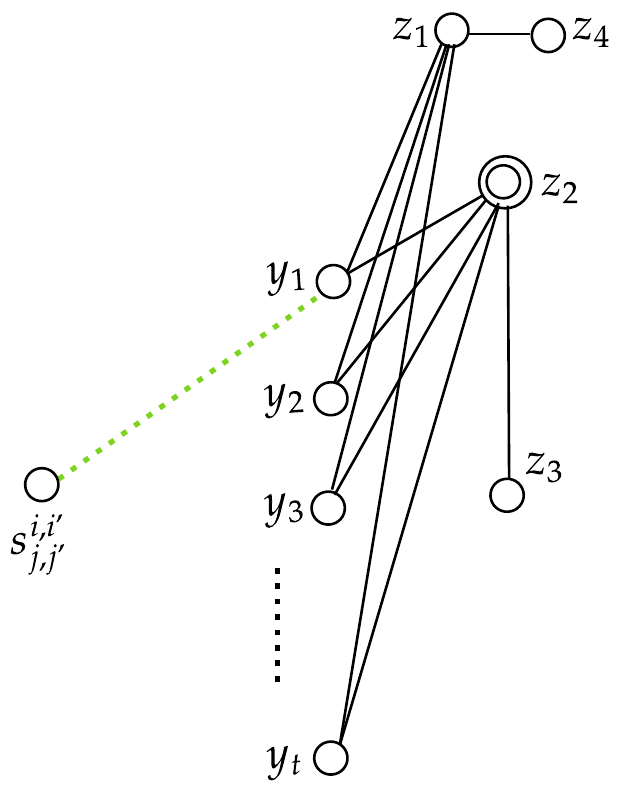}
\caption{Selector gadget}
\label{sselector}
	\end{minipage}
\end{figure}

We continue the construction of $G$ by creating a \textit{connector gadget} $S^{i,i'}_{j,j'}$, for every $i, i' \in [k]$ and $j,j' \in [n]$, where $i < i'$ and $j < j'$. The connector gadget $S^{i,i'}_{j,j'}$ contains seven vertices $\{a_1, a_2, \cdots, a_6$, $s^{i,i'}_{j,j'}\}$, where $a_5$ is a special vertex. We add edges $(a_1,a_2), (a_2,a_3), (a_2,s^{i,i'}_{j,j'}), (a_3,a_5),$\\$ (a_3,s^{i,i'}_{j,j'}), (a_4,s^{i,i'}_{j,j'}), (a_4,a_1), (a_4,a_5)$ and $(a_5,a_6)$ to complete the connector gadget. See Figure \ref{con}. Moreover, connect $S^{i,i'}_{j,j'}$ to the vertices $R$ and $B$, by adding an edge between $a_1$ with $B$ and $a_6$ with $R$. The connector gadget connects with grid vertices through $a_1$ and $a_2$. Connect the vertices $a_1$ and $a_2$ in $S_{j,j'}^{i,i'}$ with $p_{i',d'}$ and $p_{i,d}$ respectively, where $d, d'\in \{j,j'\}$.

Furthermore,  for all $i \in [k]$, we create a \textit{hanging gadget} $\mathcal{H}_i$ with seven vertices, $\{h_i, h'_i, h''_i, f_i,$ $ f'_i, d_i, d'_i\}$. Two sets of vertices $\{h_i, h'_i, h''_i, f_i\}$ and $\{h_i, h'_i, h''_i, f'_i\}$ form two cycles with four vertices. The vertices $f_i$ and $f'_i$ are special vertices in $\mathcal{H}_i$. Connect $d_i$ and $d'_i$ with $f_i$ and $f'_i$ respectively. See Figure \ref{hhang}. In addition to that, for every $h_i \in \mathcal{H}_i$, $1 \leq i \leq k$ connect with grid vertices $p_{i,j}$, for all $j\in [n]$.

Moreover, we add a \textit{selector gadget} $\mathcal{Y}$ with a set of vertices $Y=\{y_l|1 \leq l \leq t\}$ and a set of four vertices $z_1, \cdots, z_4 $ such that $z_2$ is a special vertex. For all $l \in [t]$, we connect $y_l$ with $z_1$ and $z_2$.  Additionally, add an edge between $z_2$ and $z_3$. See Figure \ref{sselector}. The vertex $z_4$ is connected to $R$ and $B$. Here the selector gadget $\mathcal{Y}$ intuitively shows which of the instances has a clique of size $k$.

In addition to these edges, for every $l \in [t]$, $i, i' \in [k]$, and $j, j' \in [n]$, add an edge $(s^{i,i'}_{j,j'}, y_l)$ if and only if $(v_j,v_{j'}) \in E(X_l)$.

Now, we have completed the construction of graph $G$. The following lemmata show the properties of the vertices in different gadgets and also help to prove the correctness of the reduction.

\begin{lemma}
\label{hanging}
For any $2$-\sh~$C$ of $G$, $C(h_i), C(h'_i)$ and $C(h''_i)$ will be $c_0$ for all values of $i \in [k]$.
\end{lemma}
\begin{proof}

For all $i \in [k]$, the vertices $f_i$ and $f'_i$ are special vertices and they are at distance two from each other. Therefore $f_i$ and $f'_i$ receive distinct colours. Since the vertices $h_i, h'_i$, and $h''_i$ are distance at most $2$ from $f_i$ and $f'_i$, we cannot colour them using red or blue.
\qed
\end{proof}

We can observe from Lemma \ref{hanging} that we cannot dominate the vertex $h_i$ by colouring a vertex from $N[h_i] \cap \mathcal{H}_i$ and all other vertices in $\mathcal{H}_i$ are dominated by at least one of the special vertices $f'_i, f_i$.

 
%


\begin{observation}
\label{riBlue}
For all $i \in [k]$, the grid vertices $\{p_{i,j} |1 \leq j \leq n \}$ are dominated by the vertex $r_i \in \mathcal{P}1_i$. 
\end{observation}

To dominate the vertex $h_i \in \mathcal{H}_i$, we need to colour a vertex from grid vertices $\{p_{i,j} |1 \leq j \leq n\}$. Since $r_i$ is coloured using blue, we can colour exactly one vertex from the set $\{p_{i,j} |1 \leq j \leq n\}$ using red colour. 

\begin{lemma}
\label{oneRed}
For any $2$-\sh~$C$ of $G$, there exist distinct $j_1, \cdots, j_k$ such that $C(p_{i,j_i})=$ red and for every other grid vertex $\{p_{i,j_t} |1 \leq t \leq n $ and $ t \neq i\}$, $C(p_{i,j_t})=c_0$.  
\end{lemma}
\begin{proof}	
Let's begin by showing that for every $i \in [k]$, there exist exactly one $j \in [n]$ such that $C(p_{i,j})=$ red. Consider the hanging gadget, it is easy to see that every vertex except $h_i$ in $\mathcal{H}_i$ are dominated by the special vertices from $\mathcal{H}_i$ and $C(h_i)= c_0$ by Lemma \ref{hanging}. It follows that indeed $\{p_{i,j}|1 \leq j \leq n \}$ contains at least one coloured vertex. The colour of the vertex in $\{p_{i,j}|1 \leq j \leq n \}$ is red because $C(r_i) =$ blue by Observation \ref{riBlue}.  Let $p_{i,j_i}$ be that coloured vertex in $\{p_{i,j}|1 \leq j \leq n \}$.

Now, we need to show that all $j_i$ are distinct. This can be proved by showing that for each $j \in [n]$, there is exactly one $i \in [k]$ such that $C(p_{i,j_i})=$ red.
Consider the vertex $w_j$ in $\mathcal{P}2_j$, observe that $N(w_j)$ contains a blue vertex in $\mathcal{P}2_j$, it follows that $w_j$ has a unique red neighbour, and thus there is exactly one $i \in [k]$ such that $C(p_{i,j})=$ red. 
Observation \ref{riBlue} also shows that $C(p_{i,j_t})=c_0$, where $t\neq i$.\qed    
\end{proof}

Now, we can consider the connector gadget $S^{i,i'}_{j,j'}$, that we created for every $p_{i',d'}$ and $p_{i,d}$, where $d,d' \in \{j,j'\}$.
The vertex $a_6$ is connected to the vertex $R$ and $a_5$ is a special vertex, $C(a_5)=$blue. The vertices $a_3, a_4, a_6$ in $S^{i,i'}_{j,j'}$ are dominated by the vertex $a_5$ and the vertex $a_1$ is dominated by $B$. 

Now, the following observations show how to dominate the vertices $a_2$ and $s^{i,i'}_{j,j'}$ in the connector gadget.

\begin{observation}
\label{Case1}
If there exists a subset square colouring $C$ such that neither $C(p_{i,d}) =$red nor $C(p_{i',d'}) =$red, where $d, d'\in \{j,j'\}$, then one of the following is true: $(a)$ $C(a_2)=$ red or $(b)$ $C(a_3)=$ red or $(c)$ $C(s^{i,i'}_{j,j'}) =$red.
\end{observation}

\begin{figure}[tbh]
\centering
\includegraphics[width=11cm]{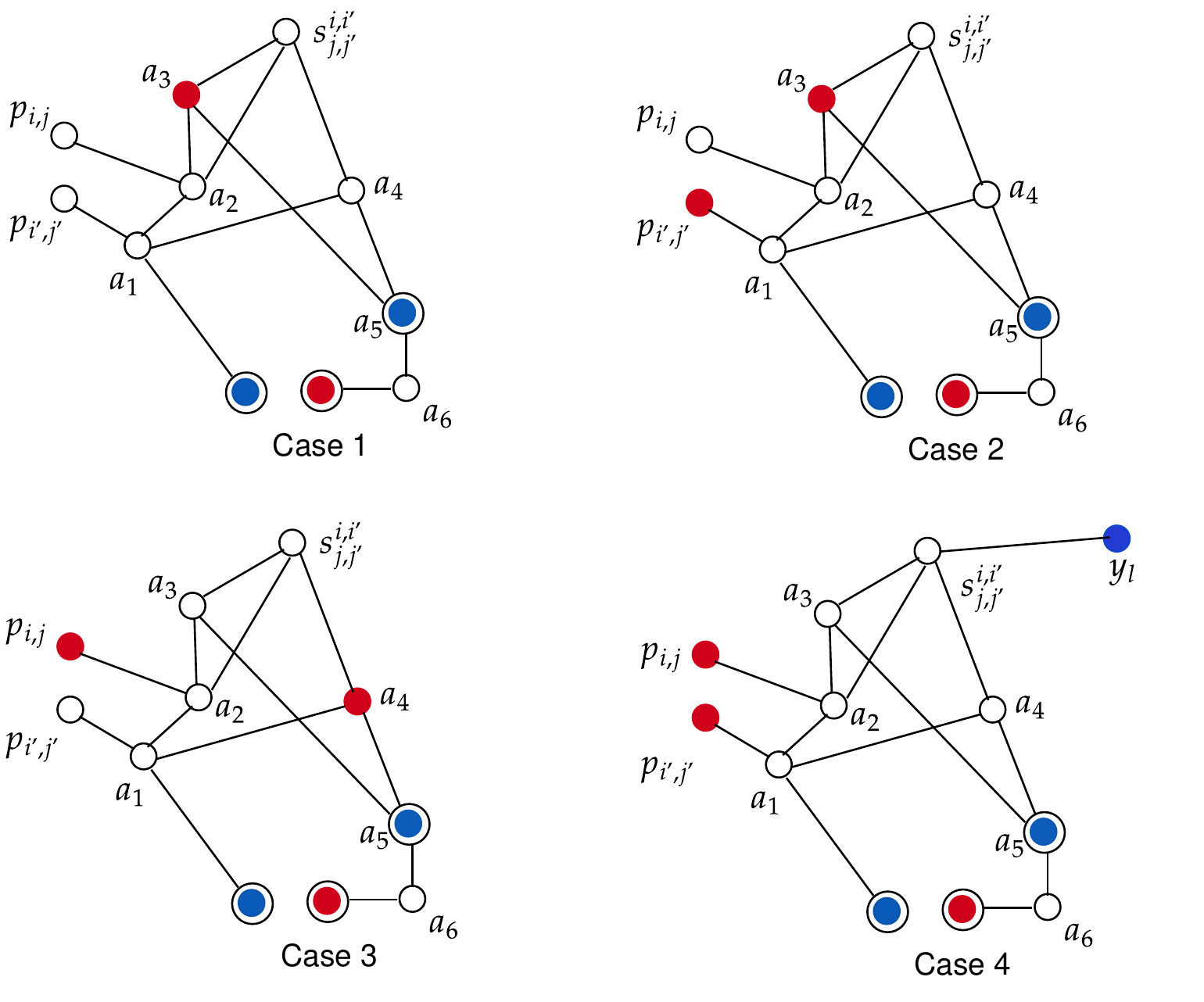}
\caption{case analysis}
\label{cases}
\end{figure}

\begin{proof}
If colour of the vertices $p_{i,d}$ and $p_{i',d'}$ is $c_0$, then to dominate the vertices $s^{i,i'}_{j,j'}$ and $a_2$ in $S^{i,i'}_{j,j'}$, we can either colour the vertex $a_3$ or $a_2$ or $s^{i,i'}_{j,j'}$. Since $C(a_5)$ is blue and $N[a_1]$ contains blue, we cannot colour any of these vertices using blue (In Figure \ref{cases}: Case 1 we coloured $a_3$). 
\end{proof}

\begin{observation}
\label{Case2}
If there exists a subset square colouring $C$ such that $C(p_{i,d})= c_0$, $C(p_{i',d'})=$ red, then one of the following is true: $(a)$ $C(a_3)= $red or $(b)$ $C(s^{i,i'}_{j,j'})=$ red.
\end{observation}

\begin{proof}
To dominate the vertices $s^{i,i'}_{j,j'}$ and $a_2$, we can either colour the vertex $a_3$ or $s^{i,i'}_{j,j'}$. Since $e(a_3,a_5)$, $e(s^{i,i'}_{j,j'},a_4)$ and $e(a_4,a_5)$ exist, we cannot colour $a_3$ or $s^{i,i'}_{j,j'}$ using blue.
\end{proof}

\begin{observation}
\label{Case3}
If there exists a subset square colouring $C$ such that $C(p_{i,d})= $ red, $C(p_{i',d'})= c_0$, then one of the following is true: $(a)$ $C(a_4)=$ red or $(b)$ the edge $(s^{i,i'}_{j,j'}, y_l)$ exists (In Figure \ref{cases}: Case 3 we coloured $a_4$) and $C(y_l)=$ blue.
\end{observation}

\begin{proof}
The vertex $a_2$ is dominated by $p_{i,d}$. Since $C(p_{i,d})$= red, we cannot colour $a_3$ or $s^{i,i'}_{j,j'}$ to dominate the vertex $s^{i,i'}_{j,j'}$. To dominate $s^{i,i'}_{j,j'}$, we can either colour $a_4$ or add an edge $(s^{i,i'}_{j,j'},y_l)$.
\end{proof}

\begin{observation}
\label{Case4}
If there exists a subset square colouring $C$ such that $C(p_{i,d})=$ red, $C(p_{i',d'})=$ red under $C$, then the edge $(s^{i,i'}_{j,j'},y_l)$ exists.
\end{observation}
\begin{proof}
If $C(p_{i,d})=$ red and $C(p_{i',d'})=$ red, then the vertex $s^{i,i'}_{j,j'}$ cannot be colour using red. Since the vertex $s^{i,i'}_{j,j'}$ is distance two from $a_5$, we cannot colour the vertex $s^{i,i'}_{j,j'}$ using blue also. Therefore edge $(s^{i,i'}_{j,j'}, y_l)$ exists to satisfy $2$-\sh. See Figure \ref{cases}: Case 4.
\end{proof}

Now, we can see that one vertex in $Y$ in the selector gadget must be coloured in order to dominate the vertex $z_1$.


\begin{lemma}
\label{oneBlue}
For any $2$-\sh~$C$ of $G$, there exists exactly one vertex in $Y$, $y_l \in \{y_i|i \in [t]\}$ such that $C(y_l)=$ blue and for every other vertex $y_i \in Y$, where $y_i \neq y_l \in Y$, $C(y_i)= c_0$.   
\end{lemma}
\begin{proof}	
The special vertex $z_2 \in \mathcal{Y}$, is distance two from $B$. Therefore $C(z_2)=$ red. Every vertex in $\mathcal{Y}$, other than $z_1$ and $z_4$ is dominated by the special vertex $z_2$. Since the vertex $z_4$ is connected to $B$ and $R$, we cannot dominate $z_1$ using $z_1$ or $z_4$. Only a vertex $\{y_l| l \in [t]\}$ can dominate the vertex $z_1$, but every $\{y_l|1 \leq l \leq t\}$ is a neighbour of $z_2$. Therefore only one vertex from $\{y_l| l \in [t]\}$ can dominate the vertex $z_1$, and the colour of that vertex will be blue. Therefore for every other vertex $y \in Y$, $C(y)= c_0$. 
\qed
\end{proof}


\noindent Now, we claim the main lemma as follows.
%

\begin{lemma}
\label{theorem2}
The graph $G$ is $2$-subset square colourable if and only if there exists $l \in [t]$ such that instance $X_l$ has a clique of size $k$.
\end{lemma} 
\begin{proof}
 
Assume there exists a YES instance $X_l$ such that it has a clique of size $k$. Let $v_{l_1}, \cdots, v_{l_k} \in V(X_l)$ be the vertices that form a clique in $X_l$. Now, for all $i \in [k], j \in [n]$, we colour the graph $G$ as follows.

\begin{enumerate}
\item Colour the vertices $R$ and $B$, such that $C(R)=$ red, and $C(B)=$ blue.

\item Let $C(f_i)=$ red and $C(f'_i)=$ blue (by Observation \ref{dum} and there exist edges $(B,d_i)$ and $(R,d'_i)$ in $G$).

\item Since $w''_j$ is connected to $R$, $C(w'_j)=$ blue.

\item $C(a_5)=$ blue (by Observation \ref{dum} and $N(N(a_5))$ contains red colour).

\item Since $z_3$ is connected to $B$, $C(z_2)=$ red.

\item Since $r'_i$ is connected to $R$, $C(r_i)=$ blue.
 
\item Let $C(y_l)=$ blue, ($y_l$ dominates the vertex $z_1$).
  
\item Let $C(p_{i,l_i})=$ red for all $i \in [k]$ (dominates the vertex $h_i$) and other vertices are coloured with $c_0$ (dominated by the vertex $r_i$) . 

\item If $C(p_{i,l_i})= c_0$ and $C(p_{i',l_{i'}})= c_0$ where $d,d'\in \{j,j'\}$, then $C(a_3)=$ red.

\item If $C(p_{i,l_i}) = c_0$ and $C(p_{i',l_{i'}})=$ red where $d,d'\in \{j,j'\}$ then $C(a_3)=$ red 

\item If $C(p_{i,l_i}) =$ red and $C(p_{i',l_{i'}})= c_0$ where $d,d'\in \{j,j'\}$, then $C(a_4)=$ red.

\end{enumerate}






\begin{figure}[tbh]
\centering
\includegraphics[width=14cm]{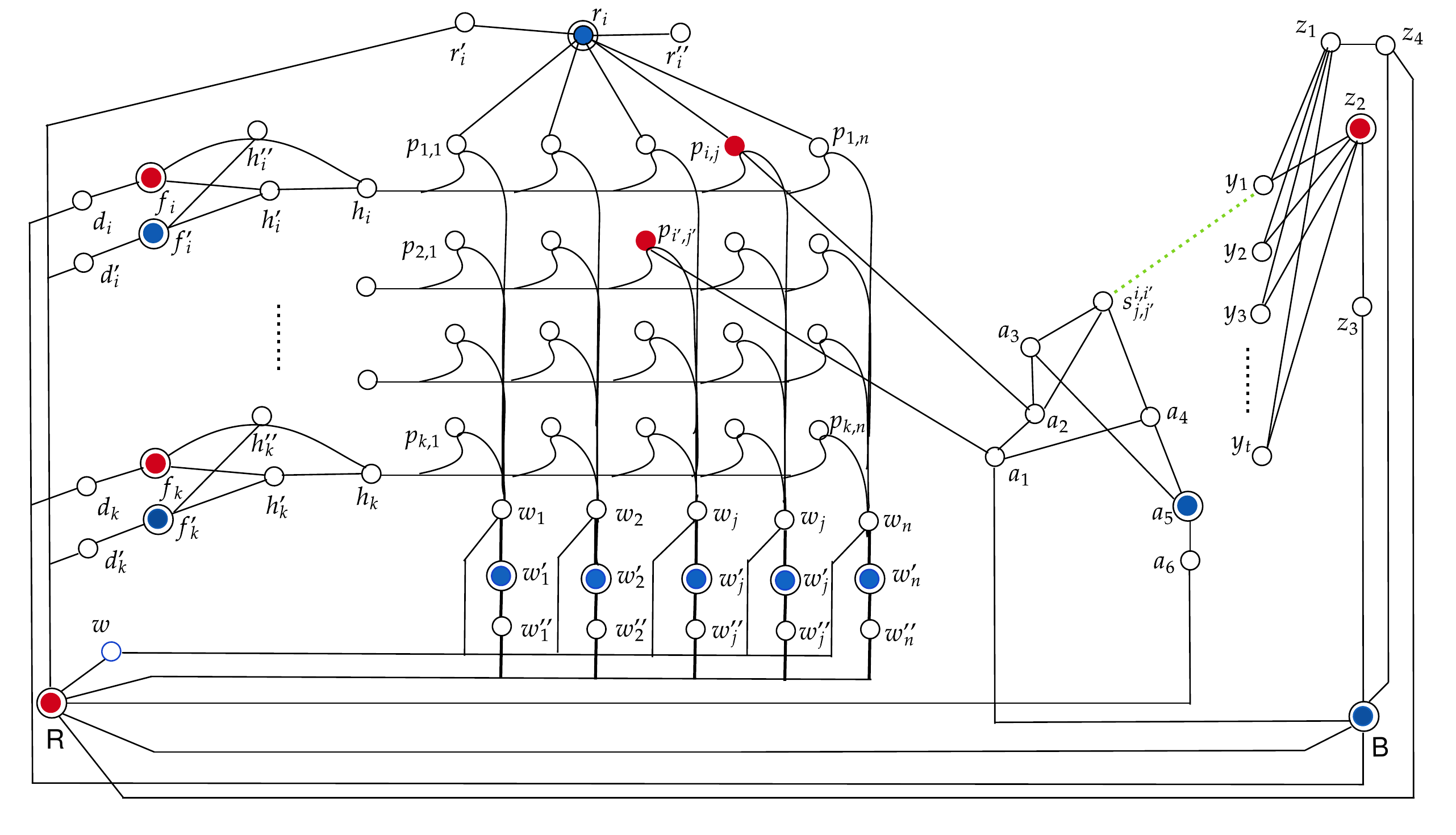}
\caption{Constructed graph $G$ with $k=4$, $n=5$ and $t=t$.}
\label{Pmain}
\end{figure}

It is easy to verify that every vertex $v$ and $N[v]$ in $G$ satisfies the $2$-\sh~constrains. Therefore this colouring in $G$ is a valid $2$-\sh. See Figure \ref{Pmain}.
 
Now we can prove if $G$ can be $2$-subset square coloured under $C$, then there exists $l \in [t]$ such that instance $X_l$ has a clique of size $k$.

From Lemma \ref{oneBlue}, we know that there exists one vertex in $Y$, with the colour blue. Let $y_l$ be that vertex. By Lemma \ref{oneRed}, there exist distinct $j_1, \cdots, j_k \in [n]$ such that $C(p_{i, j_i}) =$ red. We will show that the instance $X_l$ has a clique of size $k$, by proving that vertices $v_{j_1}, \cdots, v_{j_k} \in [n]$ form a clique in $X_l$.
Assume that this is not true. Then there exist $i, i' \in [k]$ such that $v_{j_i}$ and $v_{j_{i'}}$ are not connected by an edge in $X_l$. This leads to a contradiction. Now, consider the connector gadget $S^{i,i'}_{j_i,j'_ i}$, the vertices $p_{i',l_{i'}}$ and $p_{i,l_i}$ are coloured with red. 
Since the vertex $s^{i,i'}_{j,j'}$ is distance two from $a_5$ and $p_{i,l_i}$, we cannot colour the vertex $s^{i,i'}_{j,j'}$ using red or blue. Therefore the edge $(s^{i,i'}_{j,j'}, y_l)$ exists to dominate  $s^{i,i'}_{j,j'}$. \qed 
\end{proof}
To prove the lower bound,
it remains to bound the size of a vertex cover in $G$. Since vertices in $Y$ form an independent set in $G$, it follows that $V (G) \backslash (Y=\{y_l| 1 \leq l \leq t\})$ is a vertex cover of $G$. 

The number of grid vertices in $G$ is $nk$, and corresponding to every pair of vertices in grid there exists a connector gadget. Therefore the total number of vertices in connector gadgets is $7(n^2k^2)$. All other vertices in $G$ are linearly bounded by $n$ and $k$.
Therefore $|V (G) \backslash Y| = O(n^2 k^2)$. Thus we have proved the following theorem.

\begin{theorem}
 $2$-\sh~problem parameterized by the size of vertex cover does not have a polynomial kernel.
\end{theorem}


\subsubsection{Kernalization:  \SH~problem parameterized by the size of vertex cover, where $q \geq 3$.}
\vspace{.2cm}
\noindent In this section we extend the result for $q \geq 3$ by modifying the gadgets in $G$ and the definition of a special vertex as follows. In this section, we assume that $c_1=$ red and $c_2=$ blue.\\

\noindent \textbf{Special vertex :} If a vertex $v$ is a special vertex, then $N(v)$ contains $q+1$ single degree vertices.
\begin{observation}
\label{qdum}
For any $q$-\sh~$C$ of $G$, if a vertex $v$ in $G$ is a special vertex, then $C(v) \neq c_0$. 
\end{observation}
We add a set of special vertices $Q= \{Q_1, Q_2, \cdots Q_q\}$ such that they form a clique in $G$. Therefore all the vertices in $Q$ are coloured using different colours. Without loss of generality, assume that $C(Q_1)= c_1$, $C(Q_2)=c_2$ and so on. We modify the gadgets used in Section \ref{2nopoly} as follows.

%



\begin{figure}[th]
		\begin{minipage}[b]{0.31\linewidth}
		\centering
		\includegraphics[width=4.5cm]{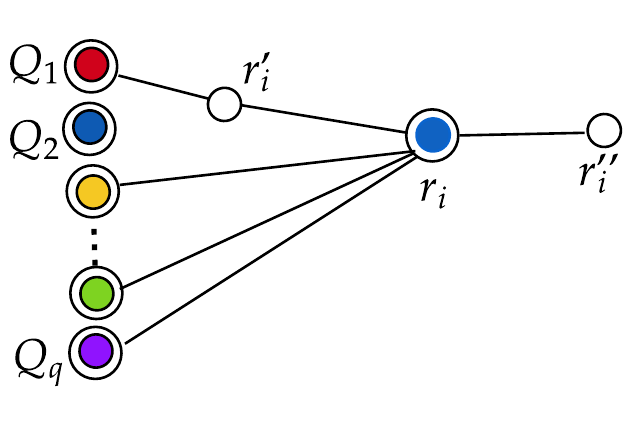}
		\caption{Path gadget $\mathcal{P}1_i$ connected with set of vertices in $Q$}
		\label{qpath}
	\end{minipage}
	\hspace{.4cm}
	\begin{minipage}[b]{0.3\linewidth}
		\centering
		\includegraphics[width=3.5cm]{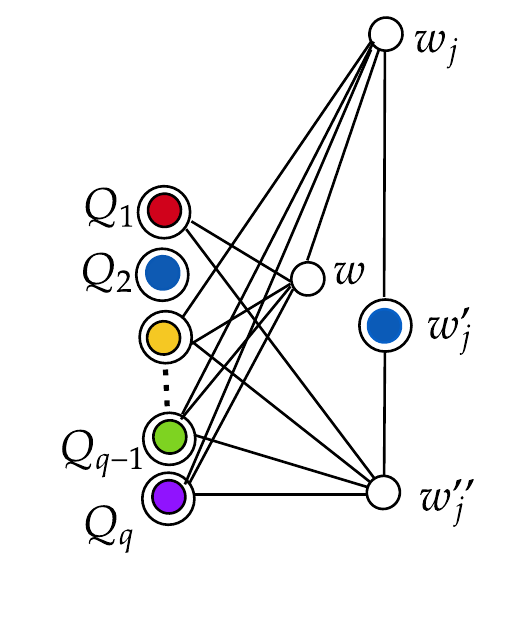}
		\caption{Path gadget $\mathcal{P}2_j$ connected with set of vertices in $Q$}
		\label{qtree}
		\end{minipage}
	\begin{minipage}[b]{0.3\linewidth}
		\centering
			\includegraphics[width=4.5cm]{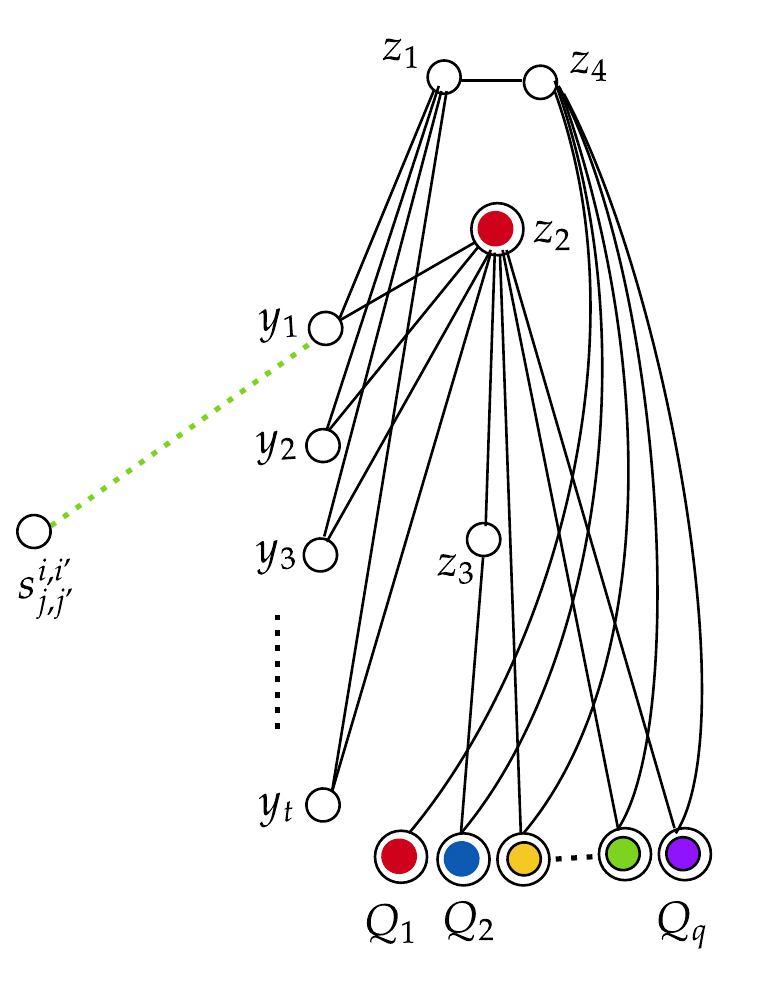}
\caption{Selector gadget}
\label{qhg}
		\end{minipage}		

\end{figure}

For every $i \in [k]$,  we add a path gadget $\mathcal{P}1_i$ with three vertices $r'_i, r_i,$ and $ r''_i$. In every $\mathcal{P}1_i$, there exist edges $(r'_i, r_i)$ and $(r_i, r''_i)$, where $r_i$ is a special vertex. We connect path gadget $\mathcal{P}1_i$ to the set of vertices in $Q$, by adding edges between $r_i$ and every vertex from the set $\{Q_3, Q_4 \cdots Q_q\}$. Also, add an edge $(r''_i, Q_1)$. See Figure \ref{qpath}. Moreover $\mathcal{P}1_i$ connects to grid vertices through $r_i$ by adding edges $(r_i, p_{i,j})$, for all $j\in [n]$. 

For every $j \in [n]$, we use a path gadget, $\mathcal{P}2_j$ ,with  three vertices $w_j$, $w'_j$ and $w''_j$. In every $\mathcal{P}2_j$, there exist edges $(w_j, w'_j)$ and $(w'_j, w''_j)$, where $w'_j$ is a special vertex. 
We add a vertex $w$ and connect path gadget $\mathcal{P}2_j$ to the set of vertices in $Q$, by adding edges between $w_j$ and every vertex from the set $\{Q_3, Q_4 \cdots Q_q\}$. Add an edge $(w_j, w)$ and add edges between $w$ and every vertex from the set $\{Q_x \in Q | 1 \leq x \leq q $ and $ x \neq 2\}$. See Figure \ref{qtree}. 

\begin{observation}
\label{qpathsSD}
For any $q$-\sh~$C$ of $G$, for all $i \in [k], j \in [n]$, $C(r_i)=$ blue and $ C(w'_j)=$ blue and all other vertices in path gadgets $\mathcal{P}1_i$ and $\mathcal{P}2_j$ are dominated by $r_i$ and $w'_j$, respectively.
\end{observation}

We modify the connector gadget as follows. For every  $i, i' \in [k]$ and $j,j' \in [n]$ in $S^{i,i'}_{j,j'}$, where $i < i'$ and $j < j'$, the gadget $S^{i,i'}_{j,j'}$ contains the set of vertices $\{a_1, a_2, a_3$, $s^{i,i'}_{j,j'}\}$, $\{a^1_4, a^2_4, \cdots a^{q-1}_4\}$, $\{a^1_5, a^2_5, \cdots a^{q-1}_5\}$, and $\{a^1_6, a^2_6, \cdots a^{q-1}_6\}$. Note that all $a^x_5$ are special vertices in $S^{i,i'}_{j,j'}$, where $1 \leq x \leq q-1$.

We add edges $(a_1,a_2), (a_2,a_3), (a_2,s^{i,i'}_{j,j'}),$ and $(a_3,s^{i,i'}_{j,j'})$ in the connector gadget. For every $1 \leq x \leq q-1$, add edges between $a^x_4$ with the vertices $s^{i,i'}_{j,j'}$ and $a_1$. In addition to that, for every $1 \leq x \leq q-1$, add edges between $a^x_4$ and $a^x_5$.
For every $1 \leq x \leq q-1$, add edges between $a^x_5$ and $a_3$. Also, for every $1 \leq x \leq q-1$, we add edges between $a^x_6$ and $a^x_5$.

\begin{figure}[th]
	\begin{minipage}[b]{0.53\linewidth}
		\centering
		\includegraphics[width=5cm]{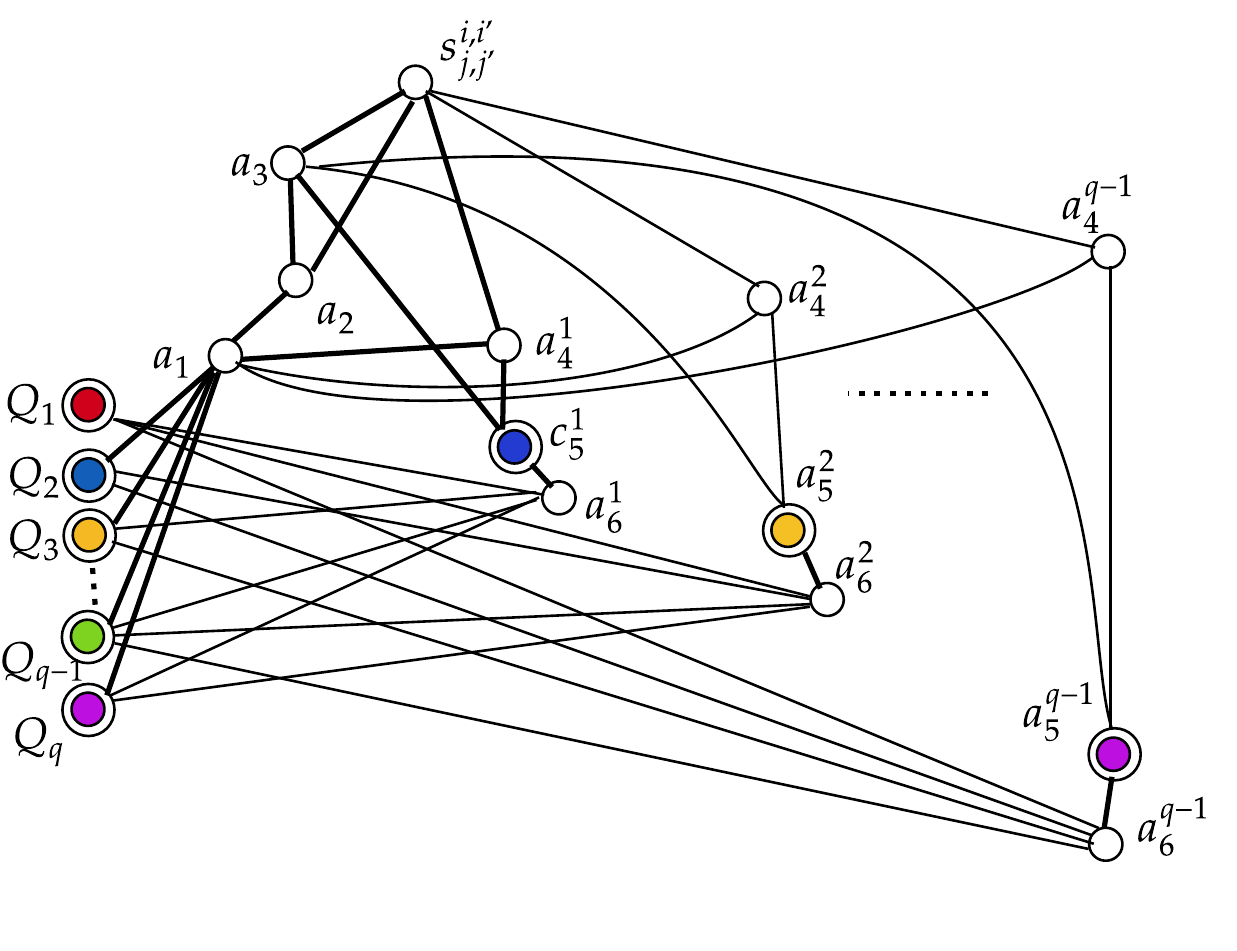}
		\caption{Connector gadget connected with a set of vertices in $Q$}
		\label{qcon}
	\end{minipage}
	\hspace{.5cm}
	\begin{minipage}[b]{0.38\linewidth}
		\centering
	\includegraphics[width=5.5cm]{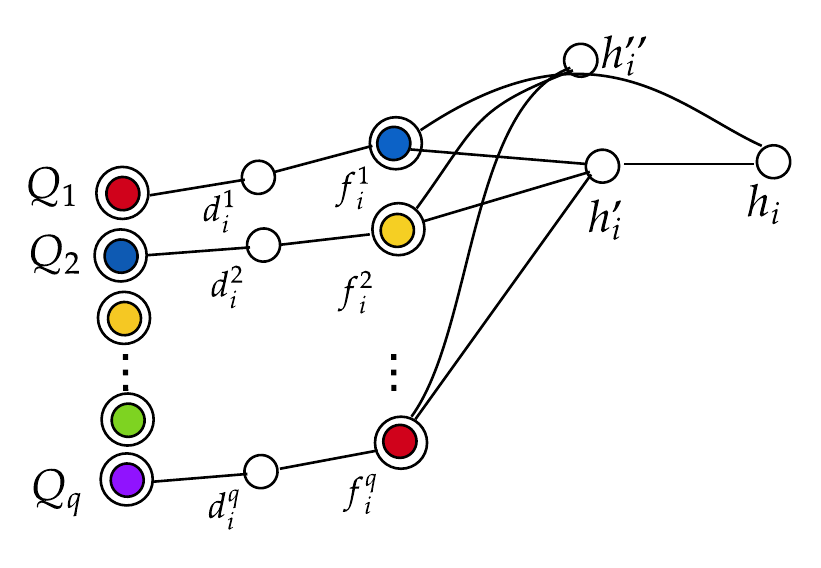}
\caption{Hanging gadget connected with a set of vertices in $Q$}
	\label{qhang}
	\end{minipage}
\end{figure}

Connect $S^{i,i'}_{j,j'}$ to the set of vertices in $Q$, by adding an edge between $a_1$ and all the vertices in $Q$ except $Q_1$. 
For every $1 \leq x \leq q-1$, we connect $a^x_6$ with all vertices in $Q$ except the vertex $Q_{x+1}$. See Figure \ref{qcon}. The connector gadget connects with grid vertices through $a_1$ and $a_2$. Connect the vertices $a_1$ and $a_2$ in $S_{j,j'}^{i,i'}$ with $p_{i',d'}$ and $p_{i,d}$ respectively, where $d, d'\in \{j,j'\}$.


Furthermore,  for all $i \in [k]$, we modify the hanging gadget as follows. In this proof the hanging gadget $\mathcal{H}_i$ contains the set of vertices, $\{h_i, h'_i, h''_i\}$, $\{f^1_i, f^2_i \cdots f^q_i\}$ and $\{d^1_i, d^2_i \cdots d^q_i\}$, where all $f^x_i$, $1 \leq x \leq q$ are special vertices in $\mathcal{H}_i$.
For all $i \in [k]$, these sets of vertices $\{h_i, h'_i, h''_i, f^x_i\}$ form $q$ cycles with four vertices, where $1 \leq x \leq q$.
For every $1 \leq x \leq q$, $d^x_i$ has an edge with $f^x_i$. The hanging gadget $\mathcal{H}_i$ is connected to the set of vertices in $Q$, by adding edges between $d^x_i$ and $Q_x \in Q$, for all $1 \leq x \leq q$. See Figure \ref{qhang}. In addition, every $h_i$ connects with grid vertices $p_{i,j}$, for all $j\in [n]$.

Furthermore, we modify the \textit{selector gadget} $\mathcal{Y}$ as follows. The selector gadget $\mathcal{Y}$, contain a set of vertices $Y=\{y_l|1 \leq l \leq t\}$ and four vertices $z_1, \cdots, z_4$, where $z_2$ is a special vertex. For all $l \in [t]$, we connect $y_l$ with $z_1$ and $z_2$. The vertex $z_4$ is connected to $z_1$. Additionally, add an edge between $z_2$ and $z_3$. 
Connect the selector gadget to the set of vertices in $Q$, by adding an edge$(z_3,Q_2)$ and by connecting the vertex $z_2$ with all $Q_x \in Q$, where $3 \leq x\leq q$. In addition to these edges, connect $z_4$ with all vertices from the set $Q$. See Figure \ref{qhg}.\\

\begin{lemma}
\label{qhanging}
For any $q$-\sh~$C$ of $G$, $C(h_i), C(h'_i)$ and $C(h''_i)$ will be $c_0$ for all values of $i \in [k]$.
\end{lemma}
\begin{proof}
For all $i \in [k]$, the vertices $f^x_i$, where $1 \leq x \leq q$ are special vertices, and they are at distance two from each other. Therefore all $f^x_i$ receive distinct $q$ colours. Since the vertices $h_i, h'_i$, and $h''_i$ are distance at most $2$ from $f^x_i$, we cannot colour them. 
\qed
\end{proof}

We can observe from Lemma \ref{qhanging} that we cannot dominate the vertex $h_i$ by colouring a vertex from $N[h_i] \cap \mathcal{H}_i$ and all other vertices in $\mathcal{H}_i$ are dominated by at least one of the special vertices $f^x_i$. Without loss of generality, assume that $C(f^x_i) = c_{i+1}$, where $1 \leq x \leq q-1$ and $C(f^x_q)= c_1$.

\begin{observation}
\label{qriBlue}
For all $i \in [k]$, the grid vertices $\{p_{i,j} |1 \leq j \leq n \}$ are dominated by the vertex $r_i \in \mathcal{P}1_i$. 
\end{observation}

To dominate the vertex $h_i \in \mathcal{H}_i$, we need to colour a vertex from grid vertices $\{p_{i,j} |1 \leq j \leq n\}$. Since $r_i$ is coloured using blue and $r_i$ is connected with all vertices in $Q$ except $Q_1$, we can colour exactly one vertex from the set $\{p_{i,j} |1 \leq j \leq n\}$ using red colour.

\begin{lemma}
\label{qoneRed}
For any $q$-\sh~$C$ of $G$, there exist distinct $j_1, \cdots j_k$ such that $C(p_{i,j_i})=$ red and for every other grid vertex $\{p_{i,j_t} |1 \leq t \leq n $ and $ t \neq i\}$, $C(p_{i,j_t})=c_0$.  
\end{lemma}
\begin{proof}	
Let's begin by showing that for every $i \in [k]$, there exists exactly one $j \in [n]$ such that $C(p_{i,j})=$ red. Consider the hanging gadget, it is easy to see that every vertex except $h_i$ in $\mathcal{H}_i$ is dominated by the special vertices from $\mathcal{H}_i$ and $C(h_i)= c_0$ by Lemma \ref{qhanging}. It follows that indeed $\{p_{i,j}|1 \leq j \leq n \}$ contains at least one coloured vertex. The colour of the vertex in $\{p_{i,j}|1 \leq j \leq n \}$ is red because $C(r_i) =$ blue and there exists edges $(r_i,Q_x)$, where $3 \leq x \leq q$ by Observation \ref{qriBlue}.  Let $p_{i,j_i}$ be that coloured vertex in $\{p_{i,j}|1 \leq j \leq n \}$.

Now, we need to show that all $j_i$ are distinct. This can be proved by showing that for each $j \in [n]$, there is at most one $i \in [k]$ such that $C(p_{i,j_i})=$ red.
Consider the vertex $w_j$ in $\mathcal{P}2_j$, observe that $N(w_j)$ contains all $q$ coloured vertices except red, it follows that $w_j$ has a unique red neighbour, and thus there is exactly one $i \in [k]$ such that $C(p_{i,j})=$ red. 
This also shows that $C(p_{i,j_t})=c_0$, where $t\neq i$. 
\end{proof}
Now we can consider the connector gadget $S^{i,i'}_{j,j'}$ that corresponds to $p_{i',d'}$ and $p_{i,d}$ where $d, d' \in \{j, j'\}$
For every $1 \leq x \leq q$, $a^x_6$ is connected to all vertices in $Q$ except $Q_{x+1}$ and all $a^x_5$ are special vertices, $C(a^x_5)= c_x$. The vertices $a_3, a^x_4, a^x_6$ in $S^{i,i'}_{j,j'}$ are dominated by the vertices $a^x_5$ and the vertex $a_1$ is dominated by $B$. 

%

Now, we can see that the colours admissible to connector gadget, selector gadget, and grid vertices are red and blue. Therefore the following lemma can be proved using arguments those given in Section \ref{2nopoly}.

Now, we can see that one vertex in $Y$ in the selector gadget must be coloured, in order to dominate the vertex $z_1$.


\begin{lemma}
\label{qoneBlue}
For any $q$-\sh~$C$ of $G$, there exists exactly one vertex in $Y$, $y_l \in \{y_i|i \in [t]\}$ such that $C(y_l)=$ blue and for every other vertex $y_i \in Y$, where $y_i \neq y_l \in Y$, $C(y_i)= c_0$.   
\end{lemma}
\begin{proof}	
The special vertex $z_2 \in \mathcal{Y}$, is distance two from all the vertices of $Q$ except $Q_1$. Therefore $C(z_2)=$ red. Every vertex in $\mathcal{Y}$, other than $z_1$ and $z_4$ are dominated by the special vertex $z_2$. Since the vertex $z_4$ is connected to all vertices in $Q$, we cannot dominate $z_1$ using $z_1$ or $z_4$. Only a vertex $\{y_l| l \in [t]\}$ can dominate the vertex $z_1$, but every $\{y_l|1 \leq l \leq t\}$ is a neighbour of $z_2$. Therefore only one vertex from $\{y_l| l \in [t]\}$ can dominate the vertex $z_1$, and the colour of that vertex will be blue. Therefore for every other vertex $y \in Y$, $C(y)= c_0$. \qed
\end{proof}
\noindent By arguments similar to those given in proof of Lemma \ref{theorem2}, we can see the following lemma.
\begin{lemma}
\label{qtheorem2}
The graph $G$ is $q$-subset square colourable if and only if there exists $l \in [t]$ such that instance $X_l$ has a clique of size $k$.
\end{lemma} 

\noindent Now, there exists a $|V (G) \backslash Y| = O(qn^2 k^2)$. Thus we have proved the following theorem.

\begin{theorem}
\SH~problem parameterized by the size of vertex cover does not have a polynomial kernel for $q \geq 3$ using  {\footnotesize OR}-cross composition from the \textsc{Clique} problem.
\end{theorem}




\subsection{Parameterized by neighbourhood diversity}
\vspace{.2cm}
\noindent We start by defining the neighbourhood diversity of a graph.
\begin{definition}
\cite{gargano2015complexity} Given a graph $G = (V,E)$, two vertices $u,v \in V$ have the same type if and only if $N(v)\backslash\{u\} = N(u)\backslash\{v\}$. The graph $G$ has a \textit{neighbourhood diversity t}, if there exists a partition of $V$ into at most $t$ sets, $V_1, V_2,\dots,V_t$ such that all the vertices in $V_i$ have the same type for $i = 1,2,\dots,t$. The family $\nu = \{V_1,V_2,\dots,V_t\}$ is called the type partition of $G$.
\end{definition}
On creating such a type partition of $V(G)$, we observe that the vertices within a partition either induces a clique or is an independent set. Further, for $1 \leq i, j \leq t$, each vertex in a partition $V_i$ is either adjacent to every vertex in another partition $V_j$ or there are no edges between vertices of $V_i$ and $V_j$.

\begin{lemma}\label{sharNDbound}
The number of colours required to subset square colour a graph $G$ is at most its neighbourhood diversity, that is, $\chi_{ssc}(G)\leq t$.
\end{lemma}
\begin{proof}
For a given graph $G$, consider its type partition $\nu$ with neighbourhood diversity $t$. Colour all isolated vertices, if any, using one of the $t$ colours. For $1 \leq i \leq t$, colour a vertex in each $V_i$ using a different colour. This gives us a subset square colouring of $V(G)$. This can be justified as follows. Vertices in partitions $V_i$ that form a clique are all dominated due to the presence of a coloured neighbour in $V_i$. 
Additionally, if vertices of $V_i$ are adjacent to partition $V_j$ for $1 \leq j \le t$, we can expect them to be adjacent to some more coloured neighbours. Now consider the case where vertices in partition $V_i$ form an independent set, then they will be adjacent to all vertices of at least one other partition $V_j$ which assures the presence of at least one coloured neighbour in the closed neighbourhood of each vertex in $V_i$. Moreover, no colour is repeated in the colouring of any connected component and we have used $t$ number of colours to subset square colour $G$. Hence the result follows.
\end{proof}
Now we show that \SH{} parameterized by $t$ is FPT by giving a polynomial kernel.




\begin{theorem}
The \SH{} problem parameterized by neighbourhood diversity admits a polynomial kernel of size $O(t^2)$.
\end{theorem}

\begin{proof}
Let $G$ be a connected graph, along with a type partition of size $t$, $t >1$. Let $G'(V',E')$ be the graph obtained from $G(V,E)$ by deleting all but $q+1$ vertices from each type partition. We will show that $G'$ can be $q$-subset square coloured if and only if $G$ can be $q$-subset square coloured. Note that $G'$ is also a connected graph.

Let $\chi$ be a $q$-subset square colouring of $G'$. We claim $\chi$ is a $q$-subset square colouring of $G$ as well. Let $V_{i}' = \{v_1, v_2, \cdots ,v_{q+1} \}$ be a vertex set in the type partition of $V(G')$. If $v_i$ and $v_j$, $1 \leq i, j \leq q+1, i \ne j$, are coloured, then $\chi(v_i) \ne \chi(v_j)$. Otherwise, there exists at least one common neighbour for all vertices in  $V_{i}'$, since $G'$ is connected and this is a contradiction. Since there are at most $q$ colours, there exists at least one uncoloured vertex, say $v_{i}$ in $V_{i}'$. Now, every vertex in $V_i \setminus V_{i}'$ is dominated in $G$ by the same vertices that dominate $v_i$ in $G'$.

In the reverse direction, assume that $G$ admits a $q$-subset square colouring, $\chi$. Then we colour the vertices $v_1, v_2, \cdots ,v_{q+1}$ in $V_{i}'$ arbitrarily using the colours, if any, used by $\chi$ on $V_{i}$. Now it is easy to see that this is a valid $q$-subset square colouring for $G'$ as well.

Now $|V(G')| \leq (q+1)t$. If $q \geq t$, the problem is trivially a YES instance, by Lemma~\ref{sharNDbound}. Therefore $|V(G')| = O(t^2)$. Now the result follows.




%
%
%
%

\end{proof}

\subsection{Parameterized by distance to cluster graph}
\vspace{.1cm}
\begin{definition}
A cluster graph is a disjoint union of complete graphs.
\end{definition}
It is easy to see that cluster graphs can be $1$-subset square coloured.
\begin{definition}
A cluster vertex deletion set $X$ of a graph $G$ is a set of vertices such that $G\backslash X$ forms a cluster graph.
\end{definition}

The cluster vertex deletion number is defined as the minimum possible size of a cluster vertex deletion set $X$. Intuitively, it is the measure of how close a graph is to being a cluster graph.

In this section, let $X$ be the subset of $V(G)$ such that $G[V\setminus X]$ is a cluster graph and $|X|=k$. Now we observe the connection between $\chi_{ssc}(G)$ and $k$.

\begin{lemma}\label{sharCVDSbound}
 $\chi_{ssc}(G)\leq |X|+1$.
\end{lemma}
\begin{proof}
To subset square colour graph $G$, we colour every vertex $v \in X$ using $k$ different colours. Now, for every complete sub-graph in $G\backslash X$, colour exactly one of the undominated vertices(if exists), using a colour that is different from previously used $k$ colours. Note that two coloured vertices $u$ and $v$ in two different complete sub-graphs do not have a common neighbour in $X$ because in such a scenario, we would not choose either of $u$ or $v$ to be coloured. We assign the $(k+1)^{th}$ colour only to a vertex not adjacent to any coloured vertex in $X$. This gives us a \sh~of $G$ as the vertices in $X$ are dominated by the colours assigned to them. Additionally, some of the vertices in $G\backslash X$ may also have a neighbour in $X$. Every $v \in V(G)$ has at least one coloured neighbour in $N[v]$ and in case more than one coloured vertices are present in $N[v]$, we are assured of them being distinct because we have used $k+1$ colours. Therefore the result follows.
\end{proof}

\begin{theorem}\label{dcg}
The \SH~is FPT when parameterized by distance to cluster graph.
\end{theorem}

\begin{proof}
If $q \geq |X|+1$, then by Lemma~\ref{sharCVDSbound} it is trivially true that a \sh~ exists. Therefore, we assume that $q \leq k$. 

If two or more vertices in a clique in $G[V\setminus X]$ have the same neighbourhood in $X$, delete all but one of those vertices. This does not affect the solution as the closed neighbourhood of the vertices are the same.
Therefore every clique in $G[V\setminus X]$ has at most $2^k$ vertices.

Now we bound the number of cliques in $G[V\setminus X]$.
Let $\{\mathcal{X}_1, \mathcal{X}_2,\dots,\mathcal{X}_{2^k}\}$ be the family of subsets of $X$. For two cliques $C_a$ and $C_b$ in $G[V\setminus X]$, we say $C_a$ and $C_b$ have the same \emph{type} if for all $1 \leq i \leq 2^k$, either both $C_a$ and $C_b$ each have a vertex whose neighbourhood in $X$ is $\mathcal{X}_i$ or neither of them has such a vertex.
Note that there can be at most $2^{2^k}$ distinct types of cliques. Now we use the following reduction exhaustively to get a reduced graph $G'$. If there exists more than $q2^k+1$ cliques of the same type, delete all but $q2^k+1$ of them. Thus there are at most $k+2^{2^k}\cdot (q2^k+1)$ vertices in $G'$.

We claim that $G$ has a $q$-subset square colouring if and only if $G'$ has a $q$-subset square colouring. Assume $G'$ admits a $q$-subset square colouring. Let $C_i$ be a clique in $G\setminus G'$. Therefore $G'$ has $(q2^k+1)$ cliques of the same type as $C_i$, let them be $C'_1,C'_2,\dots,C'_{q2^k+1}$. Similarly, for $v \in V(C_i)$, there exists $v_j \in V(C'_j)$ for all $1\leq j \leq q2^k+1$ such that $N(v) \cap X = N(v_j) \cap X$. We show that there exists at least one $C'_j$ such that one of the conditions is true.
\begin{itemize}
\item $C'_j$ contains a vertex $v_j$ such that $N(v_j) \cap X = \emptyset$ and $v_j$ is coloured.
\item None of the vertices in $C'_j$ is coloured.
\end{itemize} 
If the first condition is true, then we can dominate all vertices in $C_i$ by colouring the corresponding vertex $v$ using the same colour as $v_j$. Now, assume that the first condition is not true. Consider the vertices  $v_j \in V(C'_j)$ for all $1\leq j \leq q2^k+1$. Since all of them have common neighbours in $X$, we can colour at most $q$ such vertices. Since there are at most $2^k$ vertices in a clique, there can be at most $ q2^k$ cliques with coloured vertices. All the uncoloured vertices in a clique is dominated by vertices in $X$. The same set of vertices can dominate the vertices in $C'_j$ in $G$. The other direction is easy to see.



Since the size of the reduced instance is bounded by a function of $k$,  it follows that the \SH~is FPT when parameterized by distance to cluster graph.

\end{proof}



\subsection{Parameterized by the size of twin cover}\label{sharTwin}

\begin{definition}
\cite{twin} For a graph $G(V,E)$ a subset $X$ of vertices is a \emph{twin cover} if for every edge $e=uv \in E(G)$ either $(a)$ $u \in X$ or $v \in X$, or $(b)$ $u$ and $v$ are true twins. 
\end{definition}
Two vertices $u$ and $v$ are true twins if every other vertex is either adjacent to both $u$ and $v$ or neither of them and $u$ and $v$ has an edge between them. It follows from the definition that if $X$ is a twin cover, then $G[V\setminus X]$ is a disjoint union of cliques and for every vertex $v \in X$ and every clique in $G[V\setminus X]$, $v$ is either adjacent to every vertex in the clique or $v$ is not adjacent to any vertex in the clique. Thus for every graph $G$, distance to cluster graph of $G$ $\leq$ size of twin cover $\leq$ size of vertex cover. Thus it follows from Theorem~\ref{dcg} that the \SH{} problem parameterized by twin cover is FPT. Here, we give an algorithm with a better running time.

\begin{theorem}
The \SH~problem  parameterized by the size of twin cover is FPT.
\end{theorem}
\begin{proof}

Let $X \subset V$ be a twin cover of the graph $G(V,E)$ and $|X| =k$. It can be seen that $\chi_{ssc}(G) \leq |X|$.
If $q \geq k$, then return $TRUE$. Now, we assume that $q < k$.


Consider all possible ${(q+1)}^k$ colourings of $X$ using
$\{c_0,c_1, \dots c_q\}$. Let $\chi$ be such a colouring of $X$ such that no vertex in $V$ has the same colour repeated in its closed neighbourhood. We say a colouring $\chi_G$ of $V$ \emph{extends} $\chi$ if  $\chi_G$ restricted to $X$ is $\chi$. Our algorithm tries to extend each of the colourings to a valid $q$-subset square colouring of $G$ as follows.

In each clique in $G[V\setminus X]$, we can retain a single vertex and delete all others since all the vertices in the clique are adjacent to the same set of vertices in $X$. Now we can observe that every vertex in $G[V\setminus X]$ is independent.
 
Let $I = V\setminus X$. Let $W \subseteq I$ be the set of vertices that are not dominated by $\chi$. Note that in a valid subset square colouring of $G$ that extends $\chi$, these vertices are coloured since $I$ induces an independent set in $G$. Observe that any $v \in X$ can have at most $q$ such neighbours in $W$, because each of them needs to be coloured using a distinct colour.

The number of vertices in $W$ is therefore bounded by $kq$ which is at most $k^2$ as $q \leq k$.  We can now colour the vertices in $W$ in $\mathcal{O}(k^{k^2})$ time using brute force.
We will now be left with vertices in $X$ that are not dominated. We call the set of neighbours of such vertices in $I$ as $U$. Now if two or more vertices in $U$ have the same neighbourhood in $X$, then we retain exactly one such vertex in $U$ and delete others. This does not affect the solution since the deleted vertices are already dominated and any deleted vertex can dominate the same set of vertices as the retained vertex. Thus $U$ has at most $2^k$ vertices which we can colour in $\mathcal{O}(q^{2^k})$ or $\mathcal{O}(k^{2^k})$  time as $q \leq k$. Hence the running time of the algorithm is $\mathcal{O}(k^{2^k})$.\qed
\end{proof}




 \subsection{Parameterized by modular width}\label{sharmw}

We start by showing the connection between $\chi_{ssc}(G)$ and the modular width of the graph. We first define the modular width of a graph.
The modular width of graph $G$ is computed by virtue of four operations, namely creation of isolated vertex, disjoint union, complete join and substitution. More precisely, the modular width of $G$ equals the maximum number of operands used by any occurrence of substitution operation. These four operations that are involved in the modular decomposition of graph $G$ are described in \cite{modularwidth}. For the sake of completeness, we mention the four operations here.

\begin{definition}\label{mod_def} \cite{modularwidth} Algebraic operations involved to compute modular width of graph $G$.
\begin{itemize}
\item Create an isolated vertex;
\item The disjoint union of two graphs, i.e., the \textit{disjoint union} of two graphs $G_1$ and $G_2$, denoted by $G_1 \otimes G_2$, is the graph with vertex set $V(G_1)\cup V(G_2)$ and edge set $E(G_1)\cup E(G_2)$;
\item The complete join of two graphs, i.e., the \textit{complete join} of two graphs $G_1$ and $G_2$, denoted by $G_1\oplus G_2$, is the graph with vertex set $V(G_1)\cup V(G_2)$ and edge set $E(G_1)\cup E(G_2)\cup \{ \{v, u\}: v\in V(G_1)$ and $u\in V(G_2)\}$.
\item The substitution operation with respect to some graph $G'$ with vertices $v_1,\dots,v_n$, i.e., for graphs $G_1,\dots,G_n$ the substitution of the vertices of $G'$ by the graphs $G_1,\dots,G_n$, denoted by $G'(G_1,\dots,G_n)$, is the graph with vertex set $\underset{1 \leq i \leq n}\cup V(G_i)$ and edge set $\underset{1 \leq i \leq n} \cup E(G_i)\cup \{\{u, v\}:u \in V(G_i)$ and $v \in V(G_j)$, $v_i,v_j \in E(G')$ and $i\neq j\}$. Hence, $G'(G_1,\dots,G_n)$ is obtained from $G'$ by substituting every vertex $v_i \in V(G')$ with the graph $G_i$ and adding all edges between the vertices of a graph $G_i$ and the vertices of a graph $G_j$ whenever $\{v_i, v_j\} \in E(G')$.
\end{itemize}
\end{definition}

\begin{definition} \cite{modularwidth} Let $A$ be an algebraic expression that uses only the four operations as mentioned in Definition \ref{mod_def}. We define the width of $A$ as the maximum number of operands used by any occurrence of the substitution operation in $A$. The modular width of graph G, denoted as $w(G)$, is the least integer $m$ such that $G$ can be obtained from such an algebraic expression of width at most $m$.
\end{definition}
Cographs are the graphs that can be constructed from operations$-$creation of an isolated vertex, disjoint union of two graphs and complete join of two graphs. By definition, the modular width of cographs is two \cite{coudert2019fully}.\\

Now we state our result for the optimal algorithm as well as bounds, by bounding the number of colours used in each subgraph which is generated by one of the operations defined in \ref{mod_def}.

\begin{lemma}
\label{mww}
The maximum number of colours required to subset square colour a graph $G$ equals the modular width of $G$, that is, $\chi_{ssc}(G)\leq w(G)$.
\end{lemma}
\begin{proof}
Now we discuss the maximum number of colours required to subset square colour $G$ while we perform those operations. 
On introducing an isolated vertex, we colour it by using one colour. 
To dominate the vertices created as a result of the disjoint union of two graphs, we can use the maximum number of colours that were used in subset square colouring each subgraph in the disjoint union operation. 
Now we consider the complete join operation on the subgraph $G_c$. Let $G_c= G_{c1} \otimes G_{c2}$. If a vertex $v$ in $V(G_{ci})$, $i=\{1, 2\}$ has degree $|G_i(V)|-1$, then $\chi_{ssc}(G_c)=1$. Otherwise, we can colour an arbitrary vertex $v$ from each $G_{c1}$ and $G_{c2}$ using two distinct colours. Therefore the  value of $\chi_{ssc}$ is $ \leq 2$. 
Finally, we examine the substitution operation. We colour an arbitrary vertex $v$ in each $G_i$ using distinct colours. 
Now, for every vertex in $G_i$ is adjacent to another coloured vertex in $G_j$, where $j \neq i$. Therefore the value of $\chi_{ssc}$ can be at most $w$.  
\end{proof}


\begin{theorem}\label{FPTmw}
The \SH~is FPT when parameterized by modular width.
\end{theorem}
\begin{proof}
Let $G$ be a graph of modular width $w$. We know that $G$ can be built in a hierarchical fashion, using the operations defined in Definition~\ref{mod_def}.  Our algorithm computes the optimal subset square colouring for each intermediate subgraph created by each operation, thus eventually computing the optimal subset square colouring for $G$. Further, for each subgraph, the algorithm maintains the size of the minimum dominating set for the subgraph.

 The optimum number of colours needed to dominate a subgraph is one if the subgraph contains only isolated vertices.  The minimum dominating set of this subgraph contains all the vertices.
Now, consider the operation,  $G_d= G_{d1} \otimes G_{d2}$.  The optimum number of colours needed on $G_d$ is $max\{$optimum number of colours used on$(G_{d1})$, optimum number of colours used on$(G_{d2})\}$.  The minimum dominating set for $G_d$ is the union of minimum dominating sets of $G_{d1} $ and $G_{d2}$. 
Now, we consider a subgraph $G_c$ created by complete join operation, $G_c= G_{c1}\oplus G_{c2}$. If a vertex $v$ in $V(G_{ci})$, $i=\{1, 2\}$ has degree $|G_i(V)|-1$, then the optimum number of colours used on this node is $1$.  Otherwise, we colour one arbitrary vertex $v$ from each graph. Therefore the optimum number of colours used in this node is either $1$ or $2$.  Here, the size of the minimum dominating set is same as  the optimum number of colours used.

Now we consider the substitution operation on $G'$.  In this operation,  we replace each vertex $v_i \in V(G')$ with the subgraph $G_i$,  for $1 \leq i \leq w$. Therefore the vertices in $G'$ are in one-to-one correspondence with the subgraphs $G_i$.  Consider the optimal subset-square colouring $\chi$ of $G$. We guess the colouring $\chi '$ of $G'$ that corresponds to $\chi$ as follows : The vertex $v_i \in V(G')$ is coloured by $\chi'$ if $G_i$ contains coloured vertices under $\chi$. Moreover $\chi'(v_i)$ is one of the colours used by $\chi$ in $G_i$.  Also,  $v_i$ is not coloured by $\chi'$ if none of the vertices of $G_i$ is coloured by $\chi$.  



We can correctly guess the colouring $\chi'$ in at most $(q+1)^w$ time.  We extend the colouring $\chi'$ to $\chi$ as follows: we colour an arbitrary vertex $v$ in $G_i$,  if $v_i$ is coloured in $G' $.  Otherwise, none of the vertices in $G_i$ is coloured.  Now,  some of the vertices in $G$ are already dominated.  For a subgraph $G_j$,  if there exists an edge $(v_i, v_j) \in E(G)$ and $\chi'(v_i) \ne c_0$,  then every vertex in $G_j$ is dominated by the presence of a coloured vertex in $G_i$.  Further,  if $G_i$ contains a coloured vertex,  all the vertices in $G_i$ that are adjacent to this coloured vertex are dominated.  Thus,  a vertex $v \in V(G_i)$ is yet to be dominated if none of the neighbours of $v_i$ (in the graph $G'$) is coloured under $\chi'$,  or equivalently the corresponding subgraphs in $G$ have no coloured vertex and $v$ is not adjacent to the coloured vertex in $G_i$.  In $\chi$, such vertices in $G_i$ are dominated by colouring more vertices of $G_i$.  Note that a vertex in $G_i$ cannot be coloured using a colour already used in a subgraph $G_j$ where $v_j \in N(N(v_i))$, in $G'$. This gives us a subset of admissible colours to use in $G_i$.  Since, any pair of vertices of $G_i$ has a common neighbour, colours cannot be repeated within $G_i$.  It follows that the set of coloured vertices in such a subgraph $G_i$ corresponds to a minimum dominating set of $G_i$.  Thus the algorithm only needs to check whether the size of the minimum dominating set of $G_i$ $\leq$ the number of admissible colours of $G_i$.  It is easy to see that, in a similar way,  the minimum dominating set of $G$ can be computed starting from the minimum dominating set of $G'$, which in turn can be guessed in $O(2^w)$ time.


\noindent It follows that the \SH~is FPT when parameterized by modular width.

\end{proof}

\section{Bounds for $\chi_{ssc}$}
\label{sectionBound}
\vspace{-2pt}
\noindent
In this section, we discuss bounds on the minimum number of colours needed for subset square colouring some graph classes. 
\vspace{-5pt}
\subsection{Trees}
\vspace{-2pt}
We show that $\chi_{ssc}(G)$ can be bounded by maximum degree, when $G$ is a tree. We also show there exists a lower bound on $\chi_{ssc}(G)$ as a function of $n$.
\begin{lemma}\label{treebound}
Let $\Delta$ be the maximum degree of a tree, $T$. Then $\Delta$ colours are sufficient to subset square colour the $T$. Moreover, $\Delta-1$ colours are sometimes necessary.
\end{lemma}
\begin{proof}
Consider a tree $T$ rooted at a vertex $r$. Colour root $r$ with a colour $c_i$ where $i \neq 0$. The nodes present in its next level are left uncoloured. The nodes in following level (i.e., grandchildren of $r$) are coloured using colours $c_j$ where $j \not\in \{0,i\}$ and each of the nodes adjacent to each child of $r$ recieve a different colour. Such an assignment is accomplished due to the fact that we have $\Delta - 1$ colours after colouring $r$ with $c_i$. This preserves the characteristic of subset square colouring by assigning at most $\Delta$ unique colours to the vertices in the closed neighbourhood of each node of $T$. In other words, we colour alternate levels of a tree to subset square colour it. Therefore, either a vertex $v$ is coloured or all its neighbours are coloured using at most $\Delta$ colours.

\noindent
We remark that this may not be the optimal colouring in terms of number of colours used. However, this serves as an upper bound on the number of colours required to subset square colour a tree.

\noindent
To prove the lower bound, consider a tree $T$ of height 2. Let $r$ be the root and $d$ be the number of children of each vertex in $T$, except leaves. Therefore, maximum degree $\Delta = d + 1$. Let $v$ be a vertex at height 1. Note that to subset square colour the children of $v$, we need $d$ colours if we colour each child of $v$ using a different colour. Otherwise we have to colour $v$ in order to dominate uncoloured child node(s) of $v$. By extending this argument for each of $d - 1$ siblings of $v$, any subset square colouring of $T$ uses $d$ colours. Thus the lower bound follows. 

\end{proof}



\begin{corollary}
There exist trees that require $\Omega(\sqrt{n})$ colours to be subset square coloured.
\end{corollary}
Now, we know that Trees are a subclass of Bipartite graphs. Therefore the lower bound applies to the class of Bipartite graphs too. In the next result, we show a sub-class of Bipartite graphs can be subset square coloured using constant number of colours.
\vspace{-5pt}
\subsection{Bipartite Permutation Graphs}

In this section we discuss bounds on $\chi_{ssc}$ for bipartite permutation graphs.\\
A graph is a bipartite permutation graph if it is both bipartite and permutation graph. Let $G(A\uplus B, E)$ be a connected bipartite permutation graph, then it admits the strong ordering, adjacency and enclosure properties, as defined below~\cite{BPg}.
\begin{enumerate}
\item An ordering of the vertices $A$ in a bipartite graph $G(A\uplus B, E)$ has the \emph{adjacency property} if for each vertex $v \in B$, the vertices in $N(v)$ are consecutive in the ordering of $A$.
\item An ordering of the vertices $A$ in a bipartite graph $G(A\uplus B, E)$ has the \emph{enclosure property} if for every pair of vertices $v, u \in B$ such that $N(v)$ is a subset of $N(u)$, vertices in $N(u)-N(v)$ occur consecutively in the ordering of $A$. 
\item \emph{A strong ordering} of the vertices of a bipartite graph $G(A\uplus B, E)$ consists of an ordering of $A$ and an ordering of $B$ such that for all $(a, b')$, $(a', b)$ in $E$, where $a, a'$ are in $A$ and $b, b'$ are in $B$, $a < a'$ and $b < b'$ imply $(a, b)$ and $(a', b')$ are in $E$. 
\end{enumerate}
\begin{lemma}
If $G$ is a connected bipartite permutation graph then $\chi_{ssc}(G) \leq 4$.
\end{lemma}
Let  $A = \{a_1, a_2, \cdots a_n\}$ and $B = \{b_1, b_2, \cdots b_m\}$ have the strong ordering property. For $a_j\in A$, let $s(a_j) = \min\{i|b_i \in N(a_j)\}$ and $l(a_j) = \max\{i|b_i \in N(a_j)\}$ be the smallest and largest vertex adjacent to $a_j$ respectively. (Symmetrically defined for $B$).

Now we colour a set of vertices from $A$, such that all vertices in $B$ are dominated. In the first step, we colour the first vertex from $A$, $a_1$, using colour one.

In the $k$th step we consider the smallest $j$ such that $b_j \in B$ is not dominated. Then we colour $a_i \in A$ such that $i$ is the largest integer such that $N(a_i)$ contains $b_j$, using colour one, if $k$ is odd, or otherwise, using colour two. Repeat this till every vertex in $B$ is dominated. Now, if $a_i, a_j, a_k \in A$ are coloured in consecutive steps, then $N(a_i)$ and $N(a_k)$ are disjoint. 
For contradiction, assume that $N(a_i) \cap N(a_k) \neq \emptyset$. Let $j'=l(a_i)$. Then the vertex $b_{j'+1}$ is dominated by both $a_j$ and $a_k$. This contradicts that $a_j$ was coloured by the algorithm to dominate $b_{j'+1}$. Thus no vertex in $B$ has repeating colours in its neighborhood.

Similarly we can dominate all vertices in $A$ by colouring vertices in $B$ using colours three and four. This proves the result. \qed

Further, we show that the class of Caterpillar graphs which is a subclass of Bipartite Permutation graphs are $3$-subset square colourable. 

\begin{definition}\cite{caterpiller} 
A \emph{Caterpillar graph} is a tree such that every vertex is at distance at most one from a central path.
\end{definition}
\begin{lemma}
If $G$ is a caterpillar graph then $\chi_{ssc}(G)=3$. 
\end{lemma}
\begin{proof}
Let $P$ be the central path of $G$ with  vertices $v_1,v_2, \dots v_{n'}$. Now colouring vertices in $P$ such that $v_i$, for $1 \leq i \leq n'$, is given colour $(i\mod 3+1)$ is a valid $3$-subset square colouring. Sometimes $3$ colours are necessary to subset square colour a caterpillar graph. This can be seen by considering a caterpillar graph where every vertex in the central path is adjacent to atleast three vertices of degree one. The result follows.

\end{proof}
\vspace{-13pt}
\subsection{Threshold graph}
\begin{definition}
\cite{golumbic} A graph is a threshold graph if it can constructed from the empty graph by repeatedly adding either an isolated vertex or a dominating vertex.
\end{definition}
\begin{lemma}\label{threshold}
If $G$ is a threshold graph then $\chi_{ssc}(G) = 1$.
\end{lemma}

\begin{proof}
If isolated vertices are present, colour them using the same colour. By colouring the last introduced dominating vertex $v$, we satisfy \sh{} $G$ as each vertex in $G$ has only one coloured vertex $v$ in its closed neighbourhood.
\end{proof}
We know that the family of threshold graphs lie in the intersection of split graphs and cographs. We consider these graph classes in subsequent sections. 
\vspace{-2pt}
\subsection{Split graph}
\vspace{-1pt}
%
%
\vspace{1pt}
\begin{definition}\cite{splitD}
A graph $G$ is a split graph if $V(G)$ can be partitioned into two sets $A$ and $B$ such that $A$ induces a clique and $B$ induces an independent set.
\end{definition}
\begin{theorem}
There exist split graphs with $n$ vertices that require $\Omega(\sqrt{n})$ colours to be subset square coloured.
\end{theorem}

\begin{proof}
We will construct a split graph $G = (A\uplus B, E)$ as follows. Here $A$ induces a clique and $B$ induces an independent set. Let $A=\{ v_1, v_2, \cdots v_n\}$ and $B=\{v_{i,j}|1 \leq i,j \leq n \}$. Further we add edges from $v_{i,j} \in B$ to $v_{i} \in A$ and $v_{j} \in A$, for all $1 \leq i,j \leq n$. Note that $G$ has $n^2+n$ vertices.

All vertices $v_{i,i} \in B$ are of degree $1$. To dominate $v_{i,i} \in B$, either we need to colour $v_i$ or $v_{i,i}$, for all $1 \leq i \leq n$. If $v_i$ is coloured for all $i$, $1 \leq i \leq n$, then we need $n$ colours since all these vertices are adjacent to each other. 
Otherwise, assume there exists an $i$ such that $v_i$ is not coloured and $v_{i,i}$ is coloured. Now the $n-1$ vertices $v_{i,j} \in B$, where $i \neq j$ are dominated either by themselves or by their other neighbour $v_j \in A$. Note that here every vertex is dominated by a distinct vertex. Thus $O(n)$ vertices are coloured from at least one of the sets, $\{v_{i,j}|i \ne j\}$ and $\{v_j|j \ne i\}$. Since any two vertices from one of these sets are at distance at most $2$, $O(n)$ colours are to be used.
\end{proof}



%

\vspace{-13pt}
\subsection{Cographs}
\vspace{-1pt}
\noindent A graph with modular width two is a cograph \cite{coudert2019fully}. From Lemma \ref{mww}, the result follows. 
\begin{lemma}
\label{cogra}
If $G$ is a cograph, then $\chi_{ssc}(G) = 2$.
\end{lemma}


\begin{thebibliography}{10}

\bibitem{CFC2}
Zachary Abel, Victor Alvarez, Erik~D Demaine, S{\'a}ndor~P Fekete, Aman Gour,
  Adam Hesterberg, Phillip Keldenich, and Christian Scheffer.
\newblock Conflict-free colouring of graphs.
\newblock {\em SIAM Journal on Discrete Mathematics}, 32(4):2675--2702, 2018. 
 
\bibitem{CSR}
V.~P. Abidha, Pradeesha Ashok, Avi Tomar, and Dolly Yadav.
\newblock Coloring a dominating set without conflicts: q-subset square
  coloring.
\newblock In {\em Computer Science -- Theory and Applications}, pages 17--34.
  Springer International Publishing, 2022.	

\bibitem{ashok2022structural}
Pradeesha Ashok, Rathin Bhargava, Naman Gupta, Mohammad Khalid, and Dolly Yadav.
\newblock Structural parameterization for minimum conflict-free colouring.
\newblock In {\em Discrete Applied Mathematics}, Elsevier, 2022.

\bibitem{ashok2022polynomial}
  Pradeesha Ashok and Rajath Rao and Avi Tomar.
\newblock Polynomial Kernels for Generalized Domination Problems.
\newblock In {\em arXiv preprint arXiv:2211.03365},
  year 2022.
\bibitem{EDS}
D.~W. Bange.
\newblock Efficient dominating sets in graphs.
\newblock {\em Applications of Discrete Mathematics}, page 189–199, 1988.

\bibitem{splitD}
Alan~A Bertossi.
\newblock Dominating sets for split and bipartite graphs.
\newblock {\em Information processing letters}, 19(1):37--40, 1984.

\bibitem{5tw}
Hans~L Bodlaender, Pal~Gronass Drange, Markus~S Dregi, Fedor~V Fomin, Daniel
  Lokshtanov, and Michal Pilipczuk.
\newblock A c\^{}kn 5-approximation algorithm for treewidth.
\newblock {\em SIAM Journal on Computing}, 45(2):317--378, 2016.

\bibitem{bodlaender2014kernelization}
Hans~L Bodlaender, Bart~MP Jansen, and Stefan Kratsch.
\newblock Kernelization lower bounds by cross-composition.
\newblock {\em SIAM Journal on Discrete Mathematics}, 28(1):277--305, 2014.


\bibitem{Kloks}
Hans~L Bodlaender and Ton Kloks.
\newblock Efficient and constructive algorithms for the pathwidth and treewidth
  of graphs.
\newblock {\em Journal of Algorithms}, 21(2):358--402, 1996.


\bibitem{bodlaender2021parameterized}
Hans~L Bodlaender and  Sudeshna Kolay and Astrid Pieterse.
 \newblock Parameterized complexity of conflict-free graph coloring.
 \newblock {\em SIAM Journal on Discrete Mathematics}, 35(3), 2003--2038, 2021.

\bibitem{bu2012optimal}
Yuehua Bu and Xubo Zhu.
\newblock An optimal square colouring of planar graphs.
\newblock {\em Journal of combinatorial optimization}, 24(4):580--592, 2012.


\bibitem{Lhk}
Tiziana Calamoneri.
\newblock The l (h, k)-labelling problem: A survey and annotated bibliography.
\newblock {\em The computer journal}, 49(5):585--608, 2006.



\bibitem{chromaticBook}
Gary Chartrand and Ping Zhang.
\newblock {\em Chromatic graph theory}.
\newblock Chapman and Hall/CRC, 2008.


\bibitem{coudert2019fully}
David Coudert, Guillaume Ducoffe, and Alexandru Popa.
\newblock Fully polynomial fpt algorithms for some classes of bounded
  clique-width graphs.
\newblock {\em ACM Transactions on Algorithms (TALG)}, 15(3):1--57, 2019.
		

\bibitem{PACygan}
Marek Cygan, Fedor~V Fomin, {\L}ukasz Kowalik, Daniel Lokshtanov, D{\'a}niel
  Marx, Marcin Pilipczuk, Micha{\l} Pilipczuk, and Saket Saurabh.
\newblock {\em Parameterized algorithms}, volume~5.
\newblock Springer, 2015.

		
\bibitem{downey1999parameterized}
Rodney~G Downey, Michael~R Fellows, and Ulrike Stege.
\newblock Parameterized complexity: A framework for systematically confronting
  computational intractability.
\newblock In {\em Contemporary trends in discrete mathematics: From DIMACS and
  DIMATIA to the future}, volume~49, pages 49--99, 1999.		
		
\bibitem{erickson2010chromatic}
Lawrence~H Erickson and Steven~M LaValle.
\newblock An art gallery approach to ensuring that landmarks are
  distinguishable.
\newblock In {\em Robotics: science and systems}, volume~7, pages 81--88, 2012.


\bibitem{petr}
Ji{\v{r}}{\'\i} Fiala, Petr~A Golovach, and Jan Kratochv{\'\i}l.
\newblock Parameterized complexity of colouring problems: Treewidth versus
  vertex cover.
\newblock {\em Theoretical Computer Science}, 412(23):2513--2523, 2011.



\bibitem{modularwidth}
Jakub Gajarsk{\'y}, Michael Lampis, and Sebastian Ordyniak.
\newblock Parameterized algorithms for modular-width.
\newblock In Gregory Gutin and Stefan Szeider, editors, {\em Parameterized and
  Exact Computation}, pages 163--176, Cham, 2013. Springer International
  Publishing.

\bibitem{twin}
Robert Ganian.
\newblock Improving vertex cover as a graph parameter.
\newblock {\em Discrete Mathematics \& Theoretical Computer Science}, 17, 2015.


\bibitem{E3C}
M.~R. Garey and D.~S. Johnson.
\newblock {\em Computers and Intractability: A Guide to the Theory of
  NP-Completeness (Series of Books in the Mathematical Sciences)}.
\newblock W. H. Freeman, first edition edition, 1979.

\bibitem{gargano2015complexity}
Luisa Gargano and Adele~A. Rescigno.
\newblock Complexity of conflict-free colourings of graphs.
\newblock {\em Theoretical Computer Science}, 566:39--49, 2015.


\bibitem{golumbic}
Martin~Charles Golumbic.
\newblock Algorithmic graph theory and perfect graphs.
\newblock Elsevier.
\newblock 2004.

\bibitem{L}
Jerrold Griggs and Roger Yeh.
\newblock Labelling graphs with a condition at distance 2.
\newblock {\em SIAM J. Discrete Math.}, 5:586--595, 11 1992.

\bibitem{van2003colouring}
Jan van~den Heuvel and Sean McGuinness.
\newblock Coloring the square of a planar graph.
\newblock {\em Journal of Graph Theory}, 42(2):110--124, 2003.


\bibitem{Harmonious}
John~E Hopcroft and Mukkai~S Krishnamoorthy.
\newblock On the harmonious coloring of graphs.
\newblock {\em SIAM Journal on Algebraic Discrete Methods}, 4(3):306--311,
  1983.



\bibitem{diameter}
Daniel Lokshtanov, Neeldhara Misra, Geevarghese Philip, MS~Ramanujan, and Saket
  Saurabh.
\newblock Hardness of r-dominating set on graphs of diameter (r+ 1).
\newblock In {\em International Symposium on Parameterized and Exact
  Computation}, pages 255--267. Springer, 2013.

\bibitem{PlanarDia}
Gary MacGillivray and Karen Seyffarth.
\newblock Domination numbers of planar graphs.
\newblock {\em Journal of Graph Theory}, 22(3):213--229, 1996.

\bibitem{caterpiller}
Carmen Ortiz and M{\'o}nica Villanueva.
\newblock Maximal independent sets in caterpillar graphs.
\newblock {\em Discrete applied mathematics}, 160(3):259--266, 2012.


\bibitem{BPg}
Jeremy Spinrad, Andreas Brandstädt, and Lorna Stewart.
\newblock Bipartite permutation graphs.
\newblock {\em Discrete Applied Mathematics}, 18(3):279--292, 1987.


		
\bibitem{Yue}
Yue-Li Wang, Tsong-Wuu Lin, and Lin-Yuan Wang.
\newblock The local harmonious chromatic problem.
\newblock In {\em Proceedings of the 27th Workshop on Combinatorial Mathematics
  and Computation Theory, Taichung, Taiwan}, volume~30. Citeseer, 2010.		
		














\end{thebibliography}
\end{document}